\documentclass[12pt,preprint]{emulateapj}

\usepackage{amsmath}
\usepackage{graphicx}
\usepackage{placeins}
\usepackage{color}
\usepackage{rotating}
\usepackage{setspace}

\newcommand{\swift}{{\it Swift }}

\newcommand{\hst}{{\it HST }}
\newcommand{\ngc}{NGC\, 5548\, }

\newcommand{\psitau}{{ \psi \left( \tau | \lambda \right) }}

\newcommand{\mmdot}{{M\dot{M}}}

\newcommand{\taumean}{{\langle \tau \rangle}}

\shorttitle{AGN STORM. VI. Reverberating Disk Models for NGC 5548}
\shortauthors{}

\begin{document}

\title{Space Telescope and Optical Reverberation Mapping Project.\\
VI: Reverberating Disk Models for NGC 5548}

\author{D.~Starkey\altaffilmark{1},
Keith~Horne\altaffilmark{1},
M.~M.~Fausnaugh\altaffilmark{2},
B.~M.~Peterson\altaffilmark{2,3,7},
M.~C.~Bentz\altaffilmark{4},
C.~S.~Kochanek\altaffilmark{2,3},
K.~D.~Denney\altaffilmark{2,3},
R.~Edelson\altaffilmark{5},
M.~R.~Goad\altaffilmark{6},
G.~De~Rosa\altaffilmark{2,3,7},
M.~D.~Anderson\altaffilmark{4},
P.~Ar\'{e}valo\altaffilmark{8},
A.~J.~Barth\altaffilmark{9},
C.~Bazhaw\altaffilmark{4},
G.~A.~Borman\altaffilmark{10},
T.~A.~Boroson\altaffilmark{11},
M.~C.~Bottorff\altaffilmark{12},
W.~N.~Brandt\altaffilmark{13,14,15},
A.~A.~Breeveld\altaffilmark{16},
E.~M.~Cackett\altaffilmark{18},
M.~T.~Carini\altaffilmark{19},
K.~V.~Croxall\altaffilmark{2,3},
D.~M.~Crenshaw\altaffilmark{4},
E.~Dalla~Bont\`{a}\altaffilmark{20,21},
A.~De~Lorenzo-C\'{a}ceres\altaffilmark{1},
M.~Dietrich\altaffilmark{22,23},
N.~V.~Efimova\altaffilmark{24},
J.~Ely\altaffilmark{7},
P.~A.~Evans\altaffilmark{6},
A.~V.~Filippenko\altaffilmark{25} ,
K.~Flatland\altaffilmark{26},
N.~Gehrels\altaffilmark{27},
S.~Geier\altaffilmark{28,29,30},
J.~M.~Gelbord\altaffilmark{31,32},
L.~Gonzalez\altaffilmark{26},
V.~Gorjian\altaffilmark{33},
C.~J.~Grier,\altaffilmark{2,13,14},
D.~Grupe\altaffilmark{34},
P.~B.~Hall\altaffilmark{35},
S.~Hicks\altaffilmark{19}, 
D.~Horenstein\altaffilmark{4},
T.~Hutchison\altaffilmark{12},
M.~Im\altaffilmark{36},
J.~J.~Jensen\altaffilmark{37},
M.~D.~Joner\altaffilmark{38},
J.~Jones\altaffilmark{4},
J.~Kaastra\altaffilmark{39,40,41},
S.~Kaspi\altaffilmark{42,43},
B.~C.~Kelly\altaffilmark{44},
J.~A.~Kennea\altaffilmark{11},
S.~C.~Kim\altaffilmark{45},
M.~Kim\altaffilmark{45},
S.~A.~Klimanov\altaffilmark{24}, 
K.~T.~Korista\altaffilmark{46},
G.~A.~Kriss\altaffilmark{7,47},
J.~C.~Lee\altaffilmark{45},
D.~C.~Leonard\altaffilmark{26},
P.~Lira\altaffilmark{48},
F.~MacInnis\altaffilmark{12},
E.~R.~Manne-Nicholas\altaffilmark{4},
S.~Mathur\altaffilmark{2,3},
I.~M.~M$^{\rm c}$Hardy\altaffilmark{49},
C.~Montouri\altaffilmark{50},
R.~Musso\altaffilmark{12},
S.~V.~Nazarov\altaffilmark{10},
R.~P.~Norris\altaffilmark{4},
J.~A.~Nousek\altaffilmark{13},
D.~N.~Okhmat\altaffilmark{10},
A.~Pancoast\altaffilmark{51,56},
J.~R.~Parks\altaffilmark{4},
L.~Pei\altaffilmark{9},
R.~W.~Pogge\altaffilmark{2,3},
J.-U.~Pott\altaffilmark{54},
S.~E.~Rafter\altaffilmark{43,55},
H.-W.~Rix\altaffilmark{54},
D.~A.~Saylor\altaffilmark{4},
J.~S.~Schimoia\altaffilmark{3,56},
K.~Schn\"{u}lle\altaffilmark{54},
S.~G.~Sergeev\altaffilmark{10},
M.~H.~Siegel\altaffilmark{11},
M.~Spencer\altaffilmark{38},
H.-I.~Sung\altaffilmark{45},
K.~G.~Teems\altaffilmark{4},
C.~S.~Turner\altaffilmark{4},
P.~Uttley\altaffilmark{57},
M.~Vestergaard\altaffilmark{37,58},
C.~Villforth\altaffilmark{59},
Y.~Weiss\altaffilmark{43},
J.-H.~Woo\altaffilmark{36},
H.~Yan\altaffilmark{60},
and S.~Young\altaffilmark{5},
W.~Zheng\altaffilmark{25},  and
Y.~Zu\altaffilmark{2,61},
}

\altaffiltext{1}{SUPA Physics and Astronomy, University of
St. Andrews, Fife, KY16 9SS Scotland, UK}
\altaffiltext{2}{Department of Astronomy, The Ohio State University,
  140 W 18th Ave, Columbus, OH 43210, USA}
  
 \altaffiltext{3}{Center for Cosmology and AstroParticle Physics, The
Ohio State University, 191 West Woodruff Ave, Columbus, OH 43210, USA}
\altaffiltext{4}{Department of Physics and Astronomy, Georgia State
University, 25 Park Place, Suite 605, Atlanta, GA 30303, USA}

\altaffiltext{5}{Department of Astronomy, University of Maryland,
College Park, MD 20742-2421, USA}

\altaffiltext{6}{University of Leicester, Department of Physics and Astronomy,
Leicester, LE1 7RH, UK}

\altaffiltext{7}{Space Telescope Science Institute, 3700 San Martin
Drive, Baltimore, MD 21218, USA}

\altaffiltext{8}{Instituto de F\'{\i}sica y Astronom\'{\i}a, Facultad
de Ciencias, Universidad de Valpara\'{\i}so, Gran Bretana N 1111,
Playa Ancha, Valpara\'{\i}so, Chile}

\altaffiltext{9}{Department of Physics and Astronomy, 4129 Frederick
Reines Hall, University of California, Irvine, CA 92697, USA}

\altaffiltext{10}{Crimean Astrophysical Observatory, P/O Nauchny,
Crimea 298409, Russia}

\altaffiltext{11}{Las Cumbres Global Telescope Network, 6740 Cortona Drive, Suite 102,
Santa Barbara, CA  93117, USA}

\altaffiltext{12}{Fountainwood Observatory, Department of Physics FJS 149,
Southwestern University, 1011 E. University Ave., Georgetown, TX 78626, USA}

\altaffiltext{13}{Department of Astronomy and Astrophysics, Eberly
College of Science, The Pennsylvania State University, 525 Davey Laboratory,
University Park, PA 16802, USA}

\altaffiltext{14}{Institute for Gravitation and the Cosmos, The
Pennsylvania State University, University Park, PA 16802, USA}

\altaffiltext{15}{Department of Physics, The Pennsylvania State
University, 104 Davey Lab, University Park, PA 16802}

\altaffiltext{16}{Mullard Space Science Laboratory, University College
London, Holmbury St. Mary, Dorking, Surrey RH5 6NT, UK}

\altaffiltext{17}{Department of Statistics, The University of
Auckland, Private Bag 92019, Auckland 1142, New Zealand}

\altaffiltext{18}{Department of Physics and Astronomy, Wayne State University,
666 W. Hancock St, Detroit, MI 48201, USA}

\altaffiltext{19}{Department of Physics and Astronomy, Western Kentucky University,
1906 College Heights Blvd \#11077, Bowling Green, KY 42101, USA}

\altaffiltext{20}{Dipartimento di Fisica e Astronomia ``G. Galilei,''
Universit\`{a} di Padova, Vicolo dell'Osservatorio 3, I-35122 Padova,
Italy}

\altaffiltext{21}{INAF-Osservatorio Astronomico di Padova, Vicolo
dell'Osservatorio 5 I-35122, Padova, Italy}

\altaffiltext{22}{Department of Physics and Astronomy, Ohio
University, Athens, OH 45701, USA}

\altaffiltext{23}{Department of Earth, Environment, and Physics, Worcester
State University, 486 Chandler Street, Worcester, MA 01602, USA}

\altaffiltext{24}{Pulkovo Observatory, 196140 St.\ Petersburg, Russia}

\altaffiltext{25}{Department of Astronomy, University of California, Berkeley, CA 94720-3411, USA}

\altaffiltext{26}{Department of Astronomy, San Diego State University, San Diego, CA
  92182-1221, USA}

\altaffiltext{27}{Astrophysics Science Division, NASA Goddard Space
Flight Center, Greenbelt, MD 20771, USA}

\altaffiltext{28}{Instituto de Astrof\'{\i}sica de Canarias, 38200 La Laguna,
Tenerife, Spain}

\altaffiltext{29}{Departamento de Astrof\'{\i}sica, Universidad de La Laguna,
E-38206 La Laguna, Tenerife, Spain}

\altaffiltext{30}{Gran Telescopio Canarias (GRANTECAN), 38205 San Crist\'{o}bal de La
Laguna, Tenerife, Spain}

\altaffiltext{31}{Spectral Sciences Inc., 4 Fourth Ave., Burlington,
MA 01803, USA}

\altaffiltext{32}{Eureka Scientific Inc., 2452 Delmer St. Suite 100,
Oakland, CA 94602, USA}

\altaffiltext{33}{MS 169-327, Jet Propulsion Laboratory, California Institute of Technology, 
4800 Oak Grove Drive, Pasadena, CA 91109, USA}

\altaffiltext{34}{Space Science Center, Morehead State University, 235
Martindale Dr., Morehead, KY 40351, USA}

\altaffiltext{35}{Department of Physics and Astronomy, York
University, Toronto, ON M3J 1P3, Canada }

\altaffiltext{36}{Astronomy Program, Department of Physics \& Astronomy, 
Seoul National University, Seoul, Republic of Korea}

\altaffiltext{37}{Dark Cosmology Centre, Niels Bohr Institute,
University of Copenhagen, Juliane Maries Vej 30, DK-2100 Copenhagen,
Denmark}

\altaffiltext{38}{Department of Physics and Astronomy, N283 ESC, Brigham Young University,
Provo, UT 84602-4360, USA}

\altaffiltext{39}{SRON Netherlands Institute for Space Research,
Sorbonnelaan 2, 3584 CA Utrecht, The Netherlands}

\altaffiltext{40}{Department of Physics and Astronomy, Univeristeit
Utrecht, P.O. Box 80000, 3508 Utrecht, The Netherlands}

\altaffiltext{41}{Leiden Observatory, Leiden University, PO Box 9513,
2300 RA Leiden, The Netherlands}

\altaffiltext{42}{School of Physics and Astronomy, Raymond and Beverly Sackler Faculty of Exact
Sciences, Tel Aviv University, Tel Aviv 69978, Israel}

\altaffiltext{43}{Physics Department, Technion, Haifa 32000, Israel}

\altaffiltext{44}{Department of Physics, University of California,
Santa Barbara, CA 93106, USA}

\altaffiltext{45}{Korea Astronomy and Space Science Institute, Republic of Korea}

\altaffiltext{46}{Department of Physics, Western Michigan University,
1120 Everett Tower, Kalamazoo, MI 49008-5252, USA}

\altaffiltext{47}{Department of Physics and Astronomy, The Johns
Hopkins University, Baltimore, MD 21218, USA}

\altaffiltext{48}{Departamento de Astronomia, Universidad de Chile,
Camino del Observatorio 1515, Santiago, Chile}

\altaffiltext{49}{University of Southampton, Highfield, Southampton,
SO17 1BJ, UK}

\altaffiltext{50}{DiSAT, Universita dell'Insubria, via Valleggio 11, 22100, Como, Italy}

\altaffiltext{51}{Harvard-Smithsonian Center for Astrophysics, 60 Garden Street, Cambridge, MA 02138, USA}

\altaffiltext{52}{Department of Physics and Institute of Theoretical
and Computational Physics, University of Crete, GR-71003 Heraklion,
Greece}

\altaffiltext{53}{IESL, Foundation for Research and Technology,
GR-71110 Heraklion, Greece}

\altaffiltext{54}{Max Planck Institut f\"{u}r Astronomie, K\"{o}nigstuhl 17,
D--69117 Heidelberg, Germany} 

\altaffiltext{55}{Department of Physics, Faculty of Natural Sciences, University of Haifa,
Haifa 31905, Israel}

\altaffiltext{56}{Instituto de F\'{\i}sica, Universidade Federal do
Rio do Sul, Campus do Vale, Porto Alegre, Brazil}

\altaffiltext{57}{Astronomical Institute `Anton Pannekoek,' University
of Amsterdam, Postbus 94249, NL-1090 GE Amsterdam, The Netherlands}

\altaffiltext{58}{Steward Observatory, University of Arizona, 933
North Cherry Avenue, Tucson, AZ 85721, USA}

\altaffiltext{59}{University of Bath, Department of Physics, Claverton Down, BA2 7AY, Bath, United Kingdom}

\altaffiltext{60}{Department of Physics and Astronomy, University of
Missouri, Columbia, MO 65211, USA}

\altaffiltext{61}{Department of Physics, Carnegie Mellon University,
5000 Forbes Avenue, Pittsburgh, PA 15213, USA}

\footnotetext[3]{NSF Postdoctoral Research Fellow}
\footnotetext[56]{Einstein Fellow}
\footnotetext[62]{Packard Fellow}

\begin{abstract}
{We conduct a multiwavelength continuum variability study of the Seyfert 1 galaxy NGC 5548 to investigate the temperature structure of its accretion disk. The 19 overlapping continuum light curves ($1158\,\mathrm{\AA}$ to $9157\,\mathrm{\AA}$) combine simultaneous \hst, \swift, and ground-based observations over a 180 day period from 2014 January to July. Light-curve variability is interpreted as the reverberation response of the accretion disk to irradiation by a central time-varying point source. Our model yields the disk inclination $i=36^\circ \pm 10^\circ$, temperature $T_1=\left( 44 \pm 6 \right) \times 10^3$\,K at 1 light day from the black hole, and a temperature-radius slope ($T \propto  r^{-\alpha}$) of $\alpha = 0.99 \pm 0.03$. We also infer the driving light curve and find that it correlates poorly with both the hard and soft X-ray light curves, suggesting that the X-rays alone may not drive the ultraviolet and optical variability over the observing period. We also decompose the light curves into bright, faint, and mean accretion-disk spectra. These spectra lie below that expected for a standard blackbody accretion disk accreting at $L/L_{\rm Edd} = 0.1$.}
\end{abstract}


\section{Introduction}
\label{section:intro}

The dominant source of radiation from active galactic nuclei (AGN) is thought to be due to a blackbody-emitting accretion disk orbiting a supermassive black hole (SMBH). The inner edge of the accretion disk is determined by the spin of the black hole, and the disk temperature declines as $T(r) \propto r^{-3/4}$ \citep{ss73} for simple thin-disk models away from the inner edge of the accretion disk. Testing models of accretion disks, and measuring their properties such as their overall size scale, the logarithmic slope of the temperature profile, or the inclination of the disk relative to the observer is an ongoing challenge.

Gravitational microlensing of multiply imaged lensed quasars \citep{wa06} probes some of these issues. Microlensing studies find that disk sizes appear to be  systematically larger than predicted by thin-disk theory but scale as expected with black hole mass \citep{mo10}. The temperature profiles are close to the predictions of thin-disk theory, but the detailed microlensing results are scattered around the $T(r) \propto r^{-3/4}$ expectation and tend to have uncertainties in the logarithmic slope that limit the precision of the test \citep{bl14,bl15,jjv14}. The few (and weak) limits on the inclination of the accretion disk favor face-on geometries as would be expected for the Type~I AGN observed as gravitational lenses \citep{Po10,bl15}. The physical origin of the source of continuum variability remains unclear, but several studies point to X-rays leading ultraviolet (UV) variability \citep{mc14,mc16,tr16}. Microlensing observations of a number of gravitationally lensed quasars constrain the X-ray emitting region to lie within approximately 10 gravitational radii ($r_g = GM_\mathrm{BH}/c^2$) of the SMBH \citep{mo12,mo13,bl14}. This has also been inferred from the X-ray variability timescales for many Type 1 AGN \citep{ka16,ut14}.

Reverberation mapping \citep[RM;][]{bl82} of accretion disks provides an alternate probe of accretion-disk structure. The continuum variations at different wavelengths are correlated and systematically shows a lag that increases with wavelength if the data are of sufficiently high quality \citep{wa97,co98,se05,ca07,sh14,ed15,fa16}. The delay arises because of the different paths taken by photons emitted from the irradiating source directly toward the observer, and photons that first travel from the source to a reprocessing site on the accretion disk before re-emission to the observer (in this work we assume the reprocessing time is negligible compared to the light-travel-time effect).

A simple model for accretion-disk variability is that a variable point source (e.g., a lamppost-like source) situated a few gravitational radii above the black hole irradiates the disk \citep{AP,ca07}. Hotter, more central parts of the disk respond to the variability ahead of the cooler regions farther out. The lamppost luminosity varies stochastically in time and photons hitting the disk surface are reprocessed into UV, optical, and infrared continuum emission with light-travel-time delays that increase with wavelength as $\langle \tau \rangle \propto \lambda^{4/3}$, reflecting the standard temperature profile, $T \propto r^{-3/4}$. Evidence for this scenario has been found \citep{ca07,sh14,ed15,li15,fa16}, with mean delays broadly increasing with wavelength according to the expected result.

The AGN Space Telescope and Optical Reverberation Mapping project (STORM) collaboration has undertaken a large-scale observing campaign of NGC 5548. This object is one of the most thoroughly studied AGN and consistently exhibits significant continuum variability \citep{se07}. Paper I of the AGN STORM series \citep{ro15} presents the light curves obtained from the \textit{Hubble Space Telescope (HST)} and uses a cross-correlation analysis to obtain the light-curve time lags across the $\mathrm{C{\,IV}}$ and $\mathrm{Ly\,\mathrm{\alpha}}$ light curves. Paper II \citep{ed15} presents optical and UV light curves from \swift and finds evidence for a $\langle \tau \rangle \propto \lambda^{4/3}$ dependence of the continuum lags. Paper III \citep{fa16} adds simultaneous ground-based light curves, determined using image-subtraction methods \citep{al99}, and analyses the light curves using both cross correlation \citep{wh94} and \texttt{JAVELIN} \citep{zu11}; Paper IV \citep{go16} discusses the unexpected drop in the $\mathrm{C{\,IV}}$, $\mathrm{Si{\,IV}}$, and $\mathrm{He{\,II}}$ light curves during the NGC 5548 observing campaign; and Paper V (Pei et al., in prep) presents an analysis of the optical spectroscopic data and measures velocity resolved lags of the H$\mathrm{\beta}$ line profile. 

In this work, we analyze 19 overlapping \textit{HST}, \textit{swift}, and ground-based continuum light curves spanning $1158-9157\,\mathrm{\AA}$ over 2014 January to July. This is the same dataset presented in Paper III with the addition of the \swift\ \textit{V}-band light curve, and the reader is referred to Paper III for details on the data-reduction process. We apply a Monte Carlo Markov Chain code, \texttt{CREAM} (\textbf{C}ontinuum \textbf{RE}procesing \textbf{A}GN \textbf{M}CMC) \citep{st15,tr16}, to model these data. \texttt{CREAM} infers a disk inclination $i$, and the product of black hole mass and accretion rate $\mmdot$, assuming the time delays arise because of the thermal reprocessing of photons emitted from a central lamppost by a thin accretion disk. \texttt{CREAM} additionally infers the shape of the driving light curve, that we can then compare to the variable X-ray emission. 

This paper is organized as follows. Section \ref{sec_theory} introduces the thermal reprocessing model and outlines the \texttt{CREAM} algorithm. In Section \ref{sec_CREAM_fits} we present the CREAM fits to the AGN STORM light curves, as well as the resulting constraints on the accretion-disk inclination and temperature-radius profile. Section \ref{sec_fluxflux} presents the \texttt{CREAM}-inferred accretion-disk spectrum and discusses the implications of this for a standard blackbody accretion disk. We conclude the paper in Section \ref{section:disco} with a summary of our key findings. Throughout the paper we adopt cosmological parameters $\Omega_m = 0.28$, $\Omega_\Lambda = 0.72$, and $H_0 = 70\, \mathrm{km}$\, $\mathrm{s}^{-1}\,\mathrm{Mpc}^{-1}$ \citep{ko11}. In particular, the luminosity distance to redshift $z = 0.0172$ is $D_L = 75$ Mpc. A black hole mass $M_\mathrm{BH} = 10^{7.51}\, \mathrm{M_\odot}$ \citep{pa14} is assumed where required.

\section{Reverberating Disk Model}
\label{sec_theory}

The $\taumean \propto \lambda^{4/3}$ delay of continuum light curves is expected for thermal reprocessing of an axial, compact variable source (lamppost) irradiating a flat, blackbody accretion disk. Our model assumes that the accretion-disk flux in the UV and optical arises from blackbody emission described by the Planck function,

\begin{equation}
\label{eq_planck}
 B_\nu \left( \lambda , T \right) = \frac{2 h c}{\lambda^3}\frac{1}{e^{hc/ \lambda k T}-1},
 \end{equation}
where $h$ and $k$ are the Planck and Boltzmann constants (respectively) and $c$ is the speed of light. The disk exhibits UV and optical variability owing to irradiation by the lamppost, whose photons strike the disk and cause the temperature in Equation \ref{eq_planck} to increase locally.

The disk temperature is described by

\begin{equation}
\begin{split}
T^4 \left( t, r, \theta \right) =\frac{3GM\dot{M}}{8\pi \sigma r^3}\left(1-\sqrt{\frac{r_{\mathrm{in}}}{r}}\right) + \\
\frac{L_{b} \left( t - \tau \left( r, \theta, i \right) \right) (1-a)h}{4{\pi}{\sigma} \left( r^2 + h^2 \right)^{3/2}},
\end{split} 
\label{eqtrprof}
\end{equation}

\noindent where $\sigma$ is the Stefan-Boltzmann constant, $L_b (t)$ is the lamppost luminosity, $\tau$ is the light-travel delay between photons emitted from the lamppost and those emitted at a disk radius $r$ and azimuthal angle $\theta$, $G$ is the gravitational constant, $a$ is the disk albedo, $M$ and $\dot{M}$ are the black hole mass and accretion rate (respectively), and $h$ is the height of the lamppost above the disk plane. We adopt $r_\mathrm{in} = 6 \, r_g$, the radius of the innermost stable circular orbit (ISCO) for a Schwarzschild black hole. An observer sees a time delay, $\tau$, between the lamppost and a point at $r$ and $\theta$ of

\begin{equation}
\label{eq_taudel}
c \, \tau \left( r, \theta \right) = \sqrt{h^2 + r^2} + h\cos i - r \cos \theta \sin i, 
\end{equation}

\noindent where $i$ is the disk inclination and $i=0^\circ$ corresponds to a face-on disk.

At large radii, the disk temperature profile is $T \propto r^{-3/4}$. Since light-travel delays scale with radius as $\taumean = r/c$, and the characteristic wavelength is related to temperature by $\lambda \propto T^{-1}$, the lag of a thin accretion disk should scale as $\taumean \propto \lambda^{4/3}$. In order to explore possible deviations from the thin-disk model, we adopt a power-law temperature profile of

\begin{equation}
\label{eqtrprofparm}
T= T_{1} \left( \frac{r_1}{r} \right)^{\alpha},
\end{equation} 

\noindent where the reference temperature at radius $r_1$ is defined to be

\begin{equation}
\label{eqtr0}
T_{1}^4 = \frac{3GM\dot{M}}{8\pi \sigma r_1^3} + \frac{h \left( 1 - a \right) L_b}{4 \pi \sigma r_1^3},
\end{equation}

\noindent and we adopt a scaling radius of $r_1 = 1$ light day. Here the thin-disk limit is $\alpha = 3/4$.

\subsection{\texttt{CREAM} Fitting Code}
\label{sec_CREAM_theory}

\texttt{CREAM} is designed to fit the lamppost model to continuum AGN light curves and infer posterior probability distributions for $T_1$, $\alpha$, $\cos i$, and the lamppost light curve $X\left( t \right)$. A full description of \texttt{CREAM}, and tests using synthetic light curves, are presented by \citet{st15}. We provide here a description of \texttt{CREAM}'s basic features. 

The driving light curve $X\left( t \right)$ is modelled as a dimensionless function normalized to a mean of $\langle X \rangle = 0$ and a variance $\langle X^2 \rangle = 1$. The continuum light curve at wavelength $\lambda$ is

\begin{equation} 
\label{eqfnu}
F_{\nu}(\lambda , t)= \bar{F}_{\nu} \left( \lambda \right)  + \Delta F_{\nu}(\lambda)  \int_0^\infty \psi \left( \tau | \lambda \right) X \left( t- \tau \right) d \tau,
\end{equation}
 
\noindent where $\psi \left( \tau | \lambda \right)$ is the response function describing the contribution of the driving light curve at earlier times, $X\left( t - \tau \right)$, to the flux at wavelength $\lambda$. The response function is normalized such that 

\begin{equation}
\label{eq_psinorm}
\int_0^\infty \psitau d \tau = 1,
\end{equation} 
 
 \noindent so that the units are carried by $\Delta F_{\nu} \left( \lambda \right)$. 

\begin{figure*}
\includegraphics[scale=1.0,angle=0,trim=0cm 0.5cm 1.0cm 0cm]{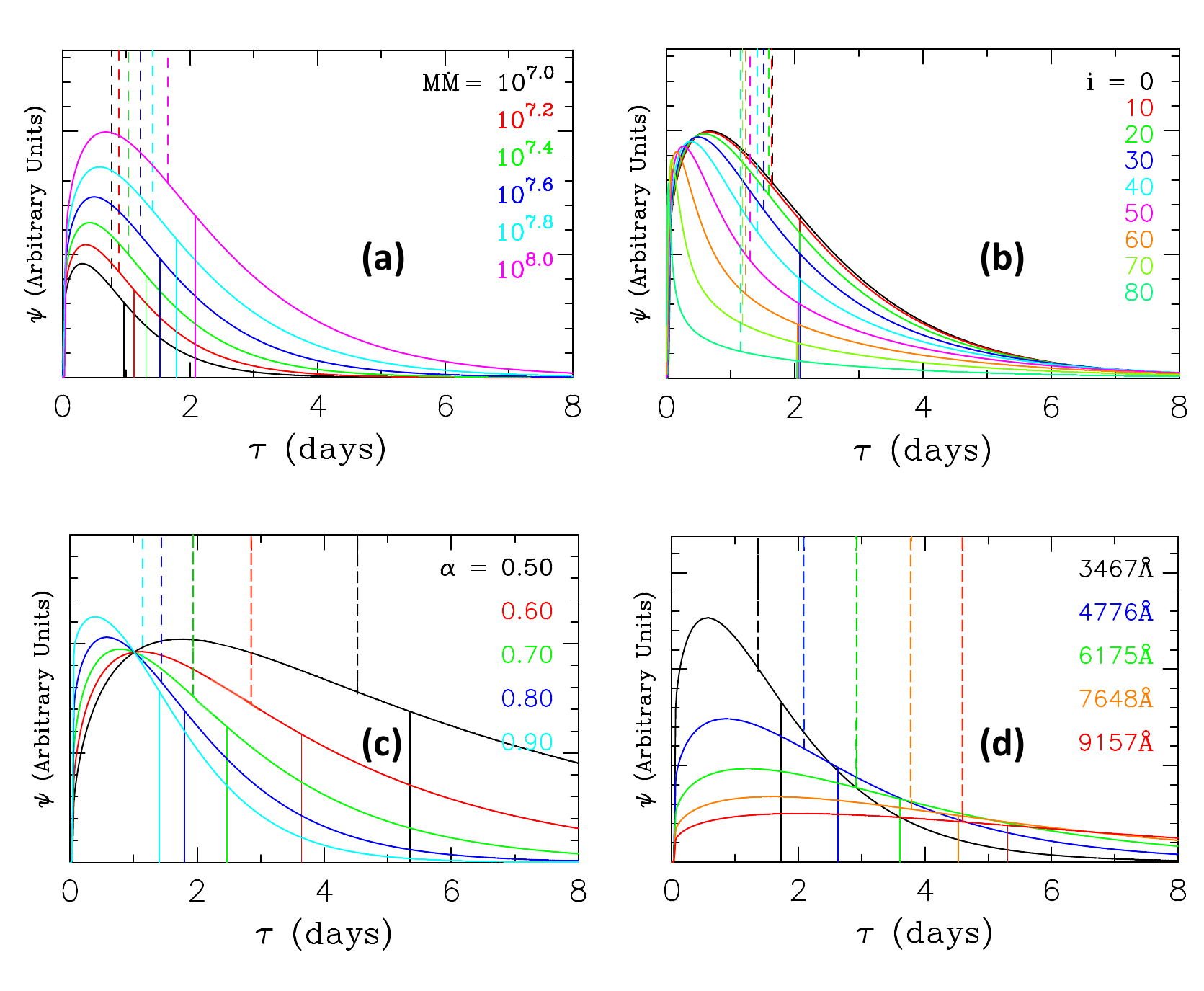}
\caption{Accretion-disk response functions for varying $\mmdot$ (a), inclination (b), temperature profile slope (c), and wavelength $\lambda$ (d). When not varied, the values are set to $\lambda = 4000\,\AA$, $i = 0$, $\alpha = 0.75$, and $\mmdot = 10^8\, \mathrm{M_\odot^2 \, yr^{-1}}$. Solid and dashed lines indicate the mean and median response function delays, respectively. Panel (b) shows that the mean delay $\langle \tau \rangle$ is inclination independent.}
\label{fig_tfdisk}
\end{figure*}

\texttt{CREAM} parametrizes $\psitau$ by $T_1$ (or equivalently $\mmdot$) and $i$. We derive the response function in the Appendix (Equation \ref{eqpsitaulam_A}); see also \citet{st15}. We show the dependence of the response function on $\lambda$, $\mmdot$, $i$, and $\alpha$ in Figure \ref{fig_tfdisk}. The response functions rise rapidly to a peak and then trail off with a long tail toward large lags. As the disk becomes edge-on, the range of time delays increases, with delays on the near side of the disk decreasing, and delays on the far side of the disk increasing relative to face-on inclinations. The effect on the response function is to skew the peak toward lower delays while increasing the long-delay tail. Solid vertical lines in Figure \ref{fig_tfdisk} show that the mean lag, $\langle \tau \rangle$, is unaffected by inclination. Increasing $\mmdot$ raises the temperature at all radii (Equation \ref{eqtrprof}), and the emission at a given wavelength arises from larger radii. Since the cooler parts of the disk are found at larger radii and emit photons at longer wavelengths, the delays increase with wavelength. The mean delays scale with $\mmdot$ and wavelength as $\left< \tau \right> \propto ( M\dot{M} ) ^{1/3} \lambda^{4/3}$.

\subsection{Driving Light Curve: $X \left( t \right)$}
\texttt{CREAM} models the driving light curve as a Fourier time series

\begin{equation}
\label{eqdrive}
X \left( t \right) = C_0 + \sum\limits_{k=1}^{N_k}C_k \cos(\omega_k t) + S_k \sin(\omega_k t),
\end{equation}

\noindent with $2N_k+1$ model parameters --- the sine and cosine amplitudes ($S_k$ and $C_k$) for each of the $N_k$ Fourier frequencies, and an offset parameter $C_0$. These driving light-curve parameters are determined as part of the fit. We use lower and upper frequencies corresponding to 300 days and 2 days (respectively), where the $k$th angular frequency $\omega_k =k \Delta \omega $ and $N_k = 150$.

\subsection{Priors}
\label{sec_priors}

Table \ref{tabprior} summarizes the model parameters and their priors. We include constant and variable components for each light curve, $\overline{F}_\nu \left( \lambda \right)$ and $\Delta F_\nu \left( \lambda \right)$. The delay distribution $\psitau$ is parameterized by $\cos i$, $T_1$, and $\alpha$. Random disk orientations are simulated using a prior uniform in $\cos i$. A uniform prior is assigned to $\alpha$, and we use log-uniform priors for the parameters $T_1$, $\overline{F}_\nu \left( \lambda \right)$ and $\Delta F_\nu \left( \lambda \right)$ to maintain positivity.

 The Fourier amplitudes control the shape of the driving light curve and require a prior to reflect the observed character of AGN light curves \citep{st15}. Without this prior, \texttt{CREAM} would assign high amplitudes to higher frequency $S_k$ and $C_k$ coefficients and overfit the data. On the timescales considered here, the driving light curve is reasonably well described by a random walk, so we assign Gaussian priors with mean 0 and variance $\sigma_k^2$, to the Fourier coefficients. The random walk is equivalent to the damped random walk (DRW) assumption from Paper III with a break timescale much larger than the observing duration. The priors take the form
 
\begin{equation}
\label{eqpriorsimp}
\sigma_k^2 = \langle S_k^2 \rangle + \langle C_k^2 \rangle = P(\omega) \Delta \omega=P_0 \Delta \omega \left( \frac{ \omega_0 }{\omega_k} \right) ^{2},
\end{equation}

\noindent where $\langle C_k^2 \rangle$ and $\langle S_k^2 \rangle$ are the mean square amplitudes of the Fourier parameters. These priors appropriately penalize high-amplitude variability on short timescales, and $P_0$ is chosen so that $\langle X^2 \rangle = 1$, 

\begin{equation}
\label{eqp0}
P_0 = \frac{2}{\omega_0^2 \Delta \omega} \left(\sum_{k=1}^{N_k} \frac{1}{\omega_{k}^2} \right)^{-1}.
\end{equation}

The light curves span 19 wavelengths $\lambda_i$ and the ground-based light curves consist of observations from multiple telescopes $N_T \left( \lambda \right)$. We incorporate the priors into a badness-of-fit (BOF) figure of merit defined by

\begin{equation}
\label{eqbof}
\begin{aligned}
\text{BOF}= \sum_{i=1}^{N_\lambda} \sum_{j=1}^{N_\mathrm{T}\left( \lambda_i \right)}  Q_{ij}^2 + 2 N_{ij} \ln \left( f_{ij} \right) \\
 + \sum_k^{N_k} \left( 2 \ln (\sigma_k^2) + \frac{C_k^2+S_k^2}{\sigma_k^2} \right).
\end{aligned}
\end{equation}

\noindent The modified $\chi^2$ term $Q_{ij}$, for $N_{ij}$ data $D_l$, model $M_l$, and errors $\sigma_l$, is

\begin{equation}
Q_\mathrm{ij}^2=\frac{1}{f_{ij}^2}\sum\limits_{l=1}^{N_\mathrm{ij}} \left( \frac{D_l-M_l}{\sigma_l} \right)^2 + \ln \sigma_l^2.
\label{eqchisq}
\end{equation}

\noindent The multiplicative factors $f_{ij}$ allow the model to adjust the nominal error bars of the light-curve points obtained at $\lambda_i$ by telescope $j$.

\begin{table}
\center
\caption{Summary of priors on each of the \texttt{CREAM} parameters.}
\begin{tabular}{ccc}
\hline
Parameter & $N_\mathrm{par}$ & Prior\\
\hline 
\hline
$S_k$ and $C_k$ & 2$N_k$ & Gaussian ($\langle S_k \rangle = \langle C_k \rangle = 0$,  \\
&& $\langle S_k^2 \rangle = \langle C_k^2 \rangle = \sigma_k^2$) \footnote{$\sigma_k$ is defined in Equation \ref{eqpriorsimp}.}  \\

$\cos i$ & 1 & Uniform\\
$\log T_1$ & 1 & Uniform \\
$\alpha$ & 1 & Uniform \\
$\log \Delta F_{\nu}$ & $N_{\lambda}$ & Uniform \\
$\log \bar{F}_{\nu}\left( \lambda \right)$ & $N_{\lambda}$ & Uniform \\
$\log f$ \footnote{$f$ is defined in Equation \ref{eqbof}.}  & $\sum_{i=1}^{N_\lambda} N_\mathrm{T} \left( i \right)$ & Uniform \\
$h \equiv 6 \mathrm{r_g}$\footnote{$h$, the lamppost height is fixed at $6r_g$ for this study.} & 1 & Uniform\\
\hline
\end{tabular}
\label{tabprior}
\end{table}

\section{\texttt{CREAM} Fits to STORM Light Curves}
\label{sec_CREAM_fits}

\begin{figure*}[H]
\centering
\includegraphics[scale=0.7,angle=0,trim=2cm 0cm 2cm 3.2cm]{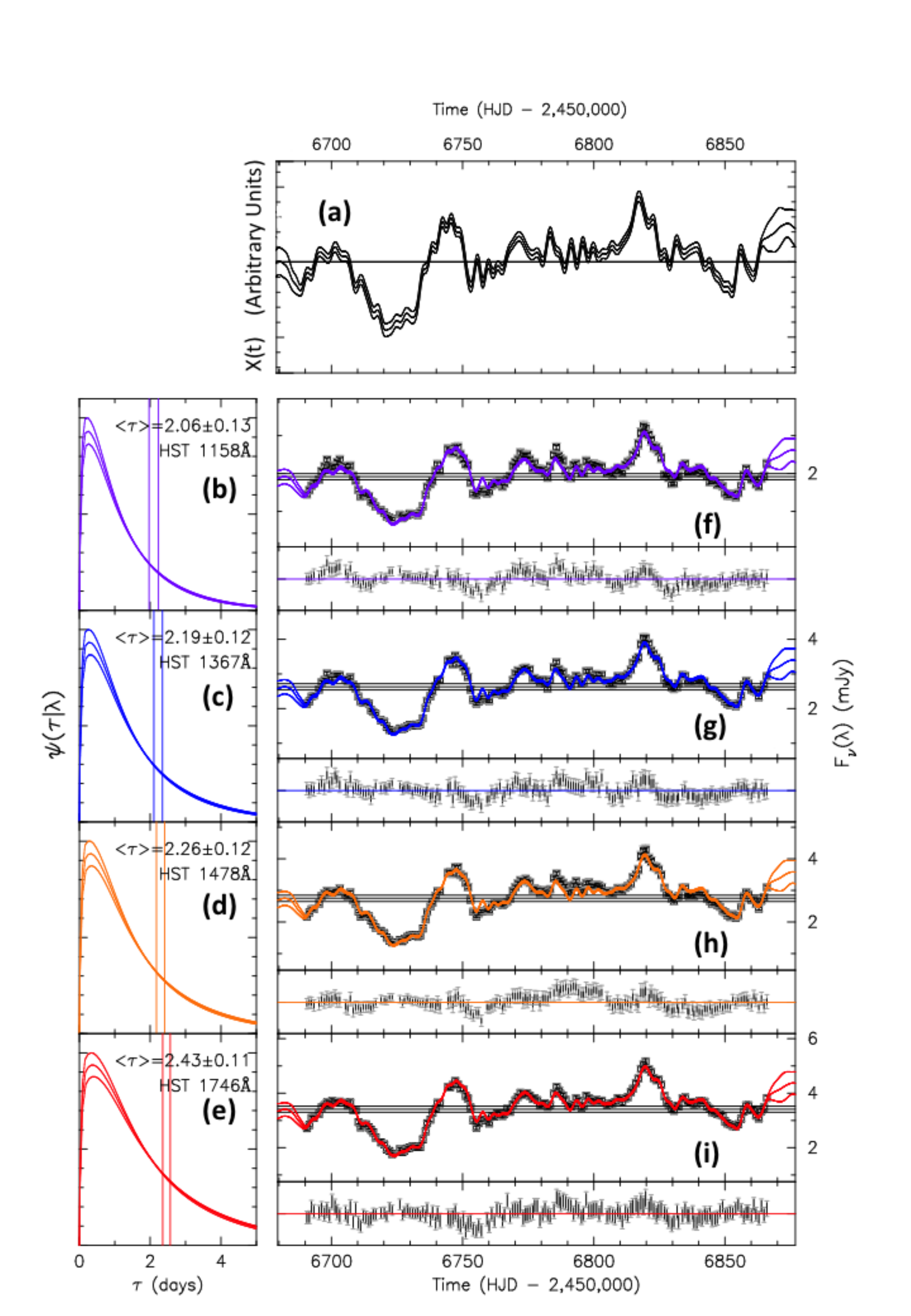}
\caption{Model 2 fits to the \textit{HST} light curves. Panels (b--e) show the mean response functions and 1$\sigma$ error envelopes from the MCMC samples. Vertical lines in these panels indicate the mean and standard deviation in $\langle \tau \rangle$. Panels (f--h) show the inferred echo light curves with residuals included beneath each light curve. Panel (a) shows the inferred driving light curve.}
\label{figcreamfit1}
\end{figure*}

\begin{figure*}[H]
\center
\includegraphics[scale=0.7,angle=0,trim=2cm 0cm 2cm 4cm]{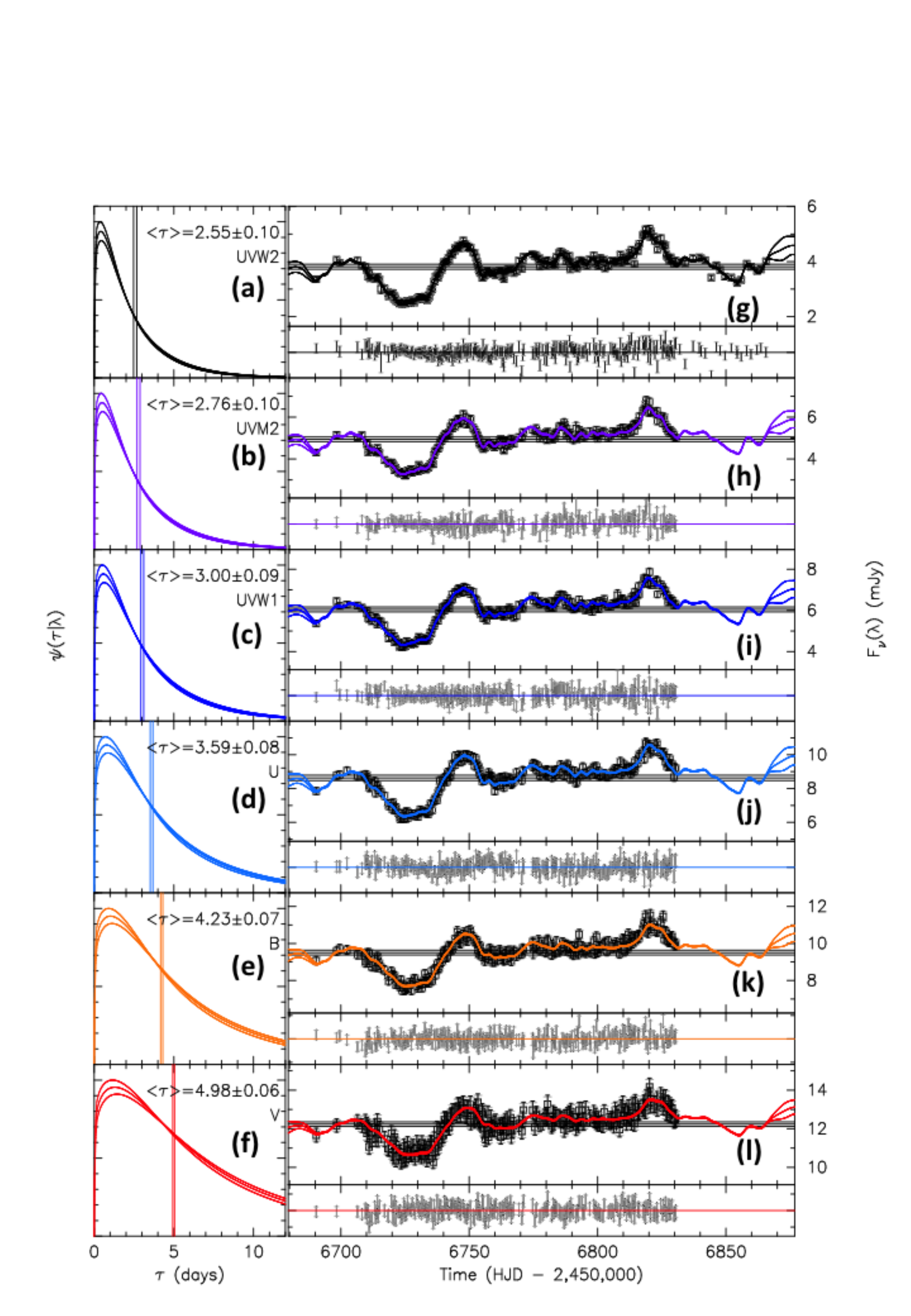}
\caption{As in Figure \ref{figcreamfit1}, but for the \swift\ light curves.}
\label{figcreamfit2}
\end{figure*}

\begin{figure*}[H]
\center
\includegraphics[scale=0.7,angle=0,trim=2cm 0cm 2cm 3cm]{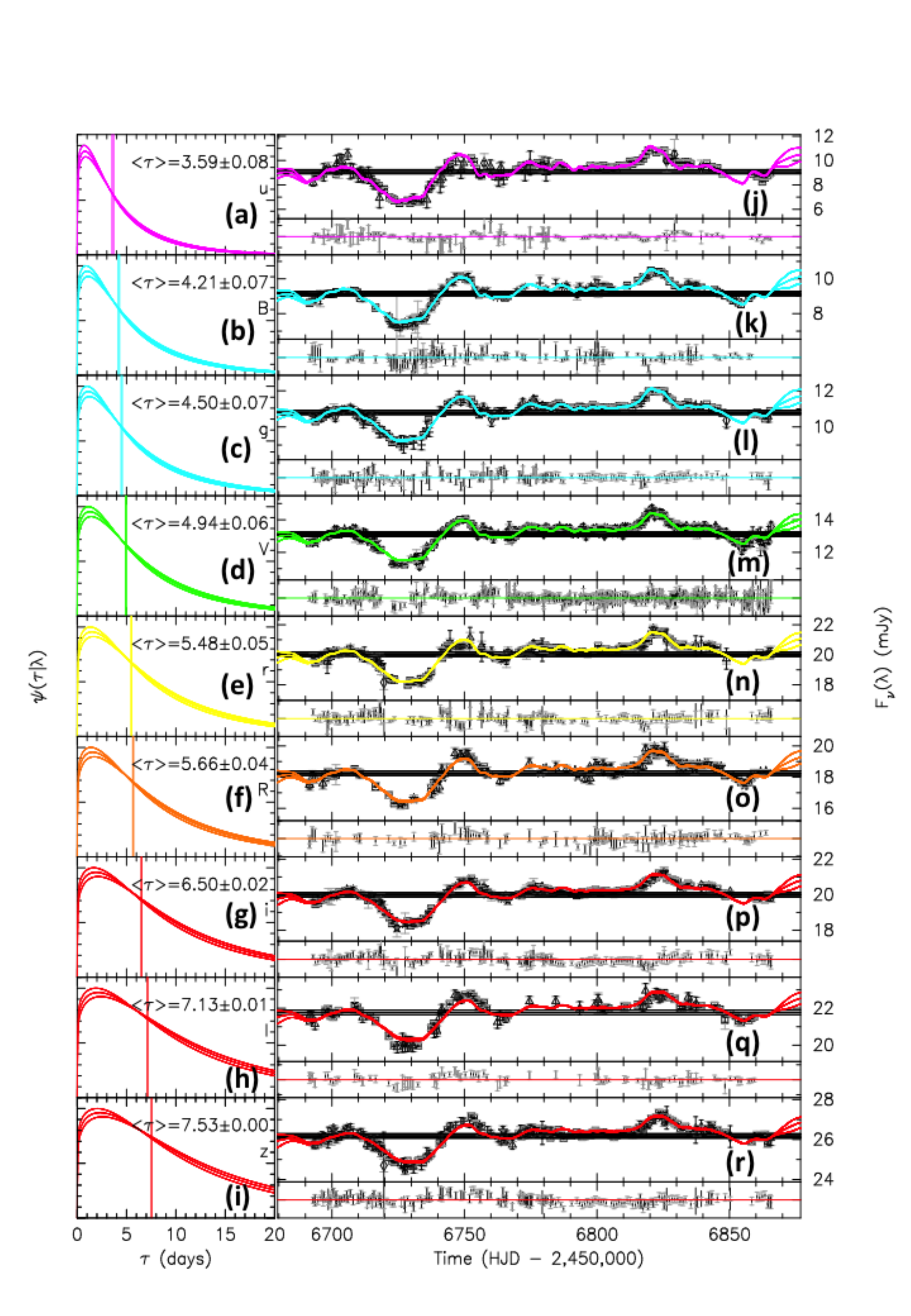}
\caption{As in Figure \ref{figcreamfit1}, but for the ground-based light curves.}
\label{figcreamfit3}
\end{figure*}

We use \texttt{CREAM} to fit the reverberating disk model to the AGN STORM light curves. We simultaneously fit all parameters in Table \ref{tabprior} except for the temperature-radius index $\alpha$ which is fixed at $\alpha \equiv 3/4$. Three independent MCMC chains, run in parallel for $10^5$ iterations, verify convergence of the parameters. Figures \ref{figcreamfit1}, \ref{figcreamfit2}, and \ref{figcreamfit3} show the fit to the \textit{HST}, \swift, and ground-based light curves, respectively. The model gives a very good fit representing all of the major as well as most of the minor features of the observed light curves. There are some significant correlated trends in the residuals for some of the bandpasses during certain time intervals. For example, the model tends to lie below the data during days 6760 and 6820 in Figure \ref{figcreamfit1} (Panels f and h). This interval lies within the period of anomalous UV and optical emission-line behaviour (see Figure 1 of Paper IV). We also note some discrepancies in the fit to the ground-based \textit{u} light curve (Figure \ref{figcreamfit3}), where the model variations seem to lead the data and have sharper features. This is probably due to contaminating Balmer continuum emission, although the \textit{u}-band error bars are relatively large owing to atmospheric telluric extinction. No significant residuals are present in the \textit{swift} fits (Figure \ref{figcreamfit2}).

\FloatBarrier

\begin{figure}
\includegraphics[scale=0.43,angle=0,trim=0cm 0cm 0cm 0cm]{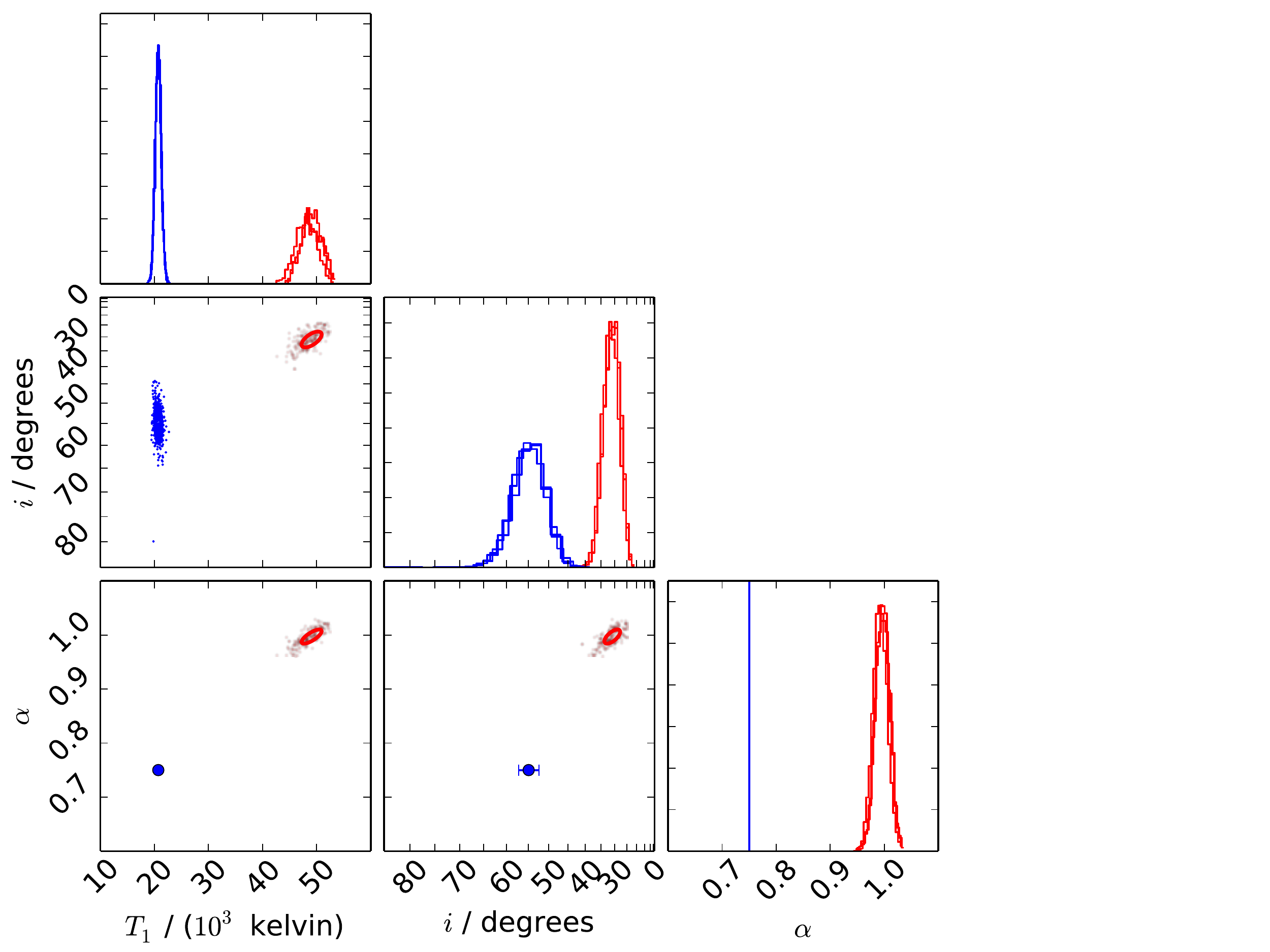}
\caption{Posterior probability histograms for the accretion-disk parameters $\alpha$, inclination, and $T_1$. Blue indicates Model 1 with $\alpha \equiv 0.75$. Red indicates Model 2 with $\alpha$ as a fitted parameter.}
\label{fig_posterior} 
\end{figure}

The posterior probability distributions for $i$, $T_1$ and $\alpha$ are shown in Figure \ref{fig_posterior} where, for $\alpha \equiv 3/4$, we find that $i=54^\circ \pm 6^\circ$ and $T_1 = \left( 22.2 \pm 0.7 \right) \times 10^3$\,K. Table \ref{tab_CREAMres} summarizes our fit results. Model 1 fits $T_1$ and $i$ with $\alpha$ fixed. Model 2 fits $T_1$, $i$, and $\alpha$.

\begin{figure}
\includegraphics[scale=0.37,angle=0,trim=1.0cm 1.0cm 0cm 0cm]{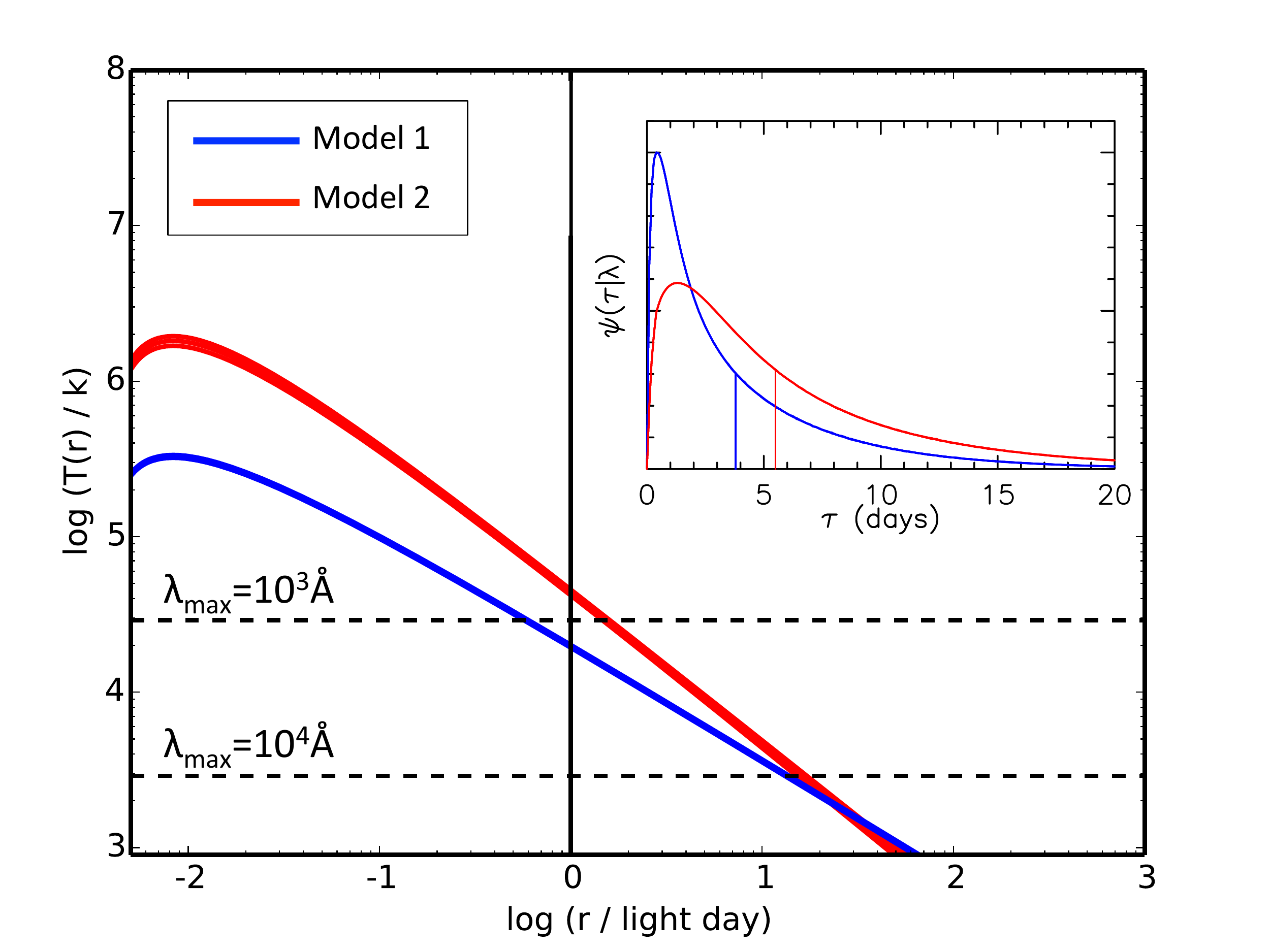}
\caption{The radial temperature profiles (Equation \ref{eqtrprofparm}) plotted for Models 1 (blue) and 2 (red) respectively. The black vertical line indicates the reference radius of 1 light day and the dashed lines mark temperatures with blackbody peak wavelengths $\lambda_\mathrm{max}$ of $10^3\,\mathrm{\AA}$ and $10^4\,\mathrm{\AA}$. The inset shows the corresponding response functions at $6000\,\mathrm{\AA}$.}
\label{fig_TvsR}
\end{figure}

Steady-state disks exhibit a $T(r)$ slope that behaves according to Equation \ref{eqtrprof}. However microlensing studies have found a range of estimated logarithmic slopes for $T(r)$ \citep{bl11,jjv14}. We therefore run a simulation allowing \texttt{CREAM} to fit the temperature-radius slope $\alpha$ (Equation \ref{eqtrprofparm}). The resulting best fits give $\alpha = 0.99 \pm 0.03$, $i=36 \pm 10^\circ$, and $T_1 = \left( 4.71 \pm 0.46 \right) \times 10^4$\,K. The resulting posterior probability distributions for $\alpha$ and $T_1$ are shown in Figure \ref{fig_posterior}. Corresponding $T(r)$ properties are shown in Figure \ref{fig_TvsR}.

\subsection{Mean Delays}

\begin{figure}
\includegraphics[scale=0.47,angle=0,trim=1cm 0cm 0cm 0cm]{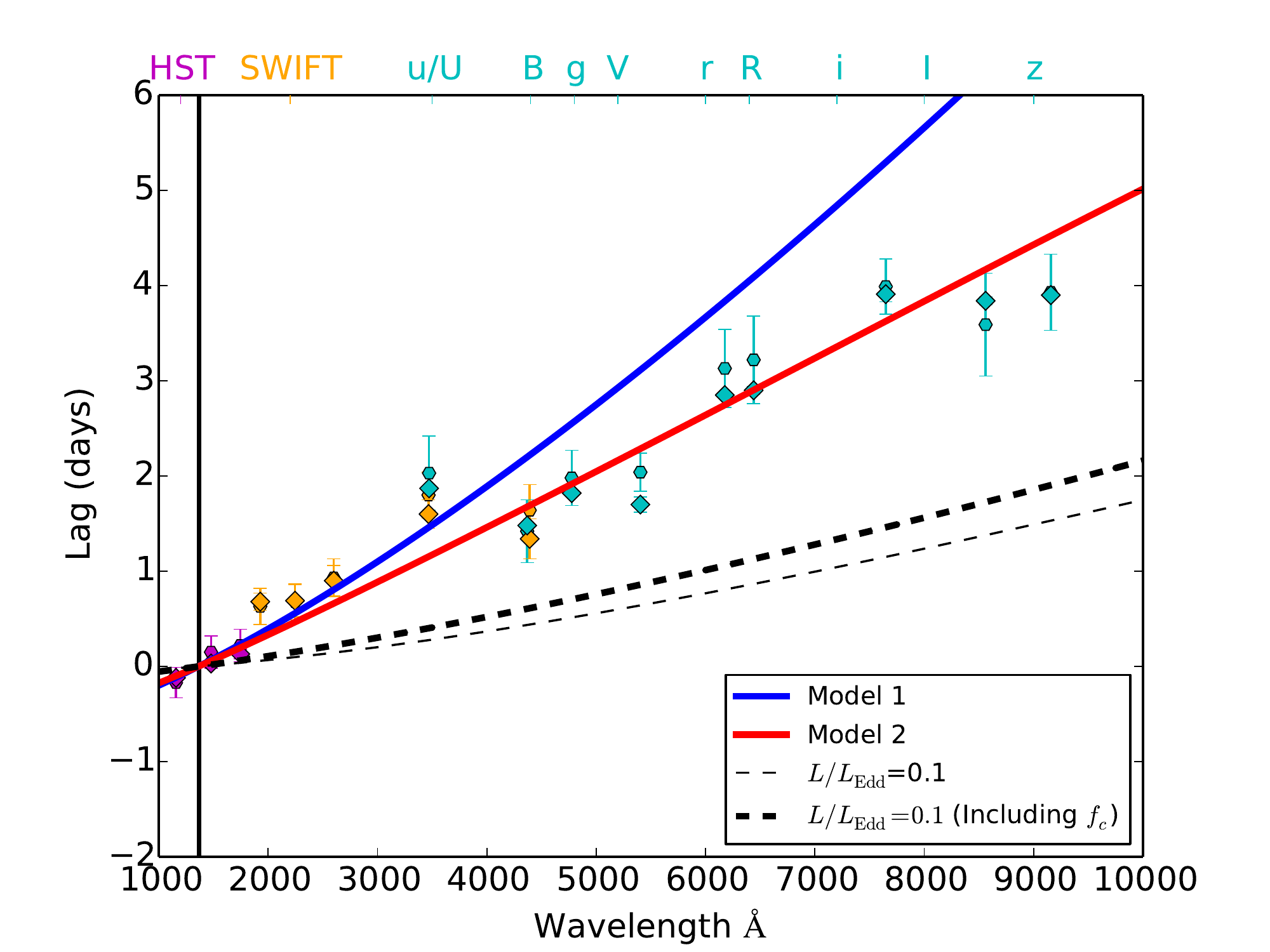}
\caption{Mean lags with $\alpha \equiv 0.75$ (Model 1, blue) and $\alpha$ fitted (Model 2, red). Markers show (circles) \texttt{Javelin} and (diamonds) CCF lags from Papers II and III for comparison. \textit{HST}, \swift and ground-based observation are coloured by magenta, orange and cyan, respectively. Lags are plotted relative to the \hst\ $1367\,\mathrm{\AA}$ light curve (vertical black line). The thin dashed line shows the lag spectrum for a standard thin disk with $L/L_\mathrm{Edd} = 0.1$. Thick dashed lines show the lag spectrum for a standard thin disk with $L/L_\mathrm{Edd} = 0.1$ that incorporates the partially covered blackbody model (Section \ref{sec_fluxflux}). } 
\label{fig_lagvswavelength}
\end{figure}

\texttt{CREAM} fits the continuum light curves directly. To produce a quantity to compare with the ICCF lags analyses of Papers II and III, we calculate the response function mean lags

\begin{equation}
\label{eq_taumean}
\langle \tau \left( \lambda \right) \rangle = \frac{\int_0^\infty \psitau \tau d\tau}{\int_0^\infty \psitau d \tau },
\end{equation}

\noindent as shown in Figure \ref{fig_lagvswavelength} alongside those inferred by \texttt{Javelin} and CCF (Paper III). We show the mean response functions and compare to the mean CCF and \texttt{Javelin} results. \citet{pe93} demonstrates that the mean of the response function is expected to agree with the mean CCF delay. 

In Paper III we fit the dependence of lag with wavelength

\begin{equation}
\label{eq_fasnaugh}
\langle \tau \rangle = \tau_0 \left[ \left( \lambda/\lambda_0 \right)^\beta - 1\right],
\end{equation}

\noindent where the $\tau_0$ term is included because the lags were measured relative to the \textit{HST} $\lambda_0 = 1367\,\mathrm{\AA}$ light curve. The index $\beta$ in the time-delay spectrum (Equation \ref{eq_fasnaugh}) corresponds to a temperature-radius slope $\alpha = 1 / \beta$ where $\alpha$ is given in Equation \ref{eqtrprofparm}. The best-fit value from Paper III ($\beta = 0.99 \pm 0.14$) agrees well with the \texttt{CREAM}-inferred value for the temperature-radius slope ($\alpha = 0.99 \pm 0.03$, thus $\beta = 1.02 \pm 0.03$). These results suggest the disk exhibits both a steeper temperature radial fall-off, and a higher disk temperature at $r_1$, than expected for a standard thin disk.

Figure \ref{fig_lagvswavelength} includes the wavelength-dependent lag spectrum for a standard $\alpha = 3/4$ disk assuming an Eddington luminosity ratio of 0.1. The lag spectrum lies above this model for both the \texttt{CREAM} and CCF analyses. We see this also in Figure \ref{fig_TvsR}, whereby $T_1$ is much larger for the red model than the blue model. 

Diffuse continuum emission (DCE) is another possibility \citep{ko01}. Here, the BLR contributes to the continuum emission as well as the disk. Time lags are proportionally larger here owing to the larger radius of the BLR than that of the disk. Paper III considers this issue and performs spectral decomposition techniques to estimate the percentage contribution of diffuse continuum emission to each light curve. This is found to be largest at \textit{u} and \textit{r} wavelengths as evidence by the mean CCF lags, that lie above the CREAM models, in Figure \ref{fig_lagvswavelength}. The DCE does not however explain the high lags, above the $L/L_\mathrm{edd}=0.1$ model, found across all wavelengths. Another possible interpretation of the large lags is that the driving light-curve photons are intercepted by some inner reprocessing medium that delays their path to the accretion disk. This inner reprocessing region is proposed by \citet{ga16}, in which the traditional accretion disk begins at closer to $200r_g$, well above the $6r_g$ value commonly thought for a non-rotating black hole. Another mechanism for truncating the accretion disk emission is a radiatively-inefficient accretion flow (RIAF) \citep{na96}. In this case, the disk itself may extend down to much smaller radii than the model of \citet{ga16}, but ceases to radiate at low radii. \citet{de11} introduce an inhomogeneous disk model in which temperature fluctuations occur randomly throughout the disk rather than being driven by a lamppost. This model successfully explains the large accretion disk sizes found from microlensing studies \citep{mo10,bl11}, but lacks detailed predictions on the lag-wavelength profile.

\subsection{Error-Bar Rescaling}
Estimates of the error bar scale factors $f_{ij}$ (Equation \ref{eqbof}) are given in Table \ref{tabsigexpand}. These scale factors are with respect to the error bars adopted in Paper III. The \texttt{CREAM} estimates for the $f$ factors are determined by a competition between the BOF $\chi^2$ term and the $2 N_j \ln f$ term that penalizes large $f$ values (Equation \ref{eqbof}). An $f$ factor greater than unity may indicate an underestimate of the error bars.

We see that the \swift points consistently yield $f$ values close to unity, indicating good agreement between the reverberating disk model and the data for these points. The fits to the \hst light curves yield $f$ factors around 2. A deviation from lamppost-model behavior can either be interpreted as the nominal error bars being too small, or of variability not adequately modeled by \texttt{CREAM}'s linearised echo model. In some cases the ground-based observations require a significant error bar rescaling factor to reconcile the model with these data (Table \ref{tabsigexpand_2}). Model 1 appears to consistently require larger rescaling factors for each telescope, and this introduces a larger penalty in the BOF (Table \ref{tab_CREAMres}).

There is also a correlation between $f_\mathrm{ij}$ and the number of data points per telescope for the ground-based data. This correlation is not seen in tests with synthetic light curves and may indicate an artefact of the calibration process. We note, however, that models of the data with $f \equiv 1$ yield comparable results for $\alpha$, $T_1$, and inclination.

\begin{table}
\center
\caption{\texttt{CREAM} model parameter inferences and fit statistics.}
\label{tab_CREAMres}
\begin{tabular}{ccc}
\hline
& Model 1 & Model 2  \\
\hline
$T_1$ ($10^4$k) & $22.2 \pm 0.7$ & $4.71 \pm 0.46$ \\

$i$ (deg) & $54 \pm 6$ & $36 \pm 10$              \\
$\alpha$  & $\equiv 0.75$ & $0.99 \pm 0.03$       \\
$\chi^2/(Nf^2)$ & $0.97$ & $0.98$                           \\
$\sum_i 2N_i\ln f_i$ \footnote{This term indicates the penalty applied for expanding the error bars summed over telescopes $i$.} & $3571$  & $3466$                                \\
\hline 
\hline
\end{tabular}
\end{table}

\subsection{The Driving Light Curve vs. X-rays}

\begin{figure*}
\center
\includegraphics[scale=0.9,angle=0,trim=1.4cm 0cm 0cm 0cm]{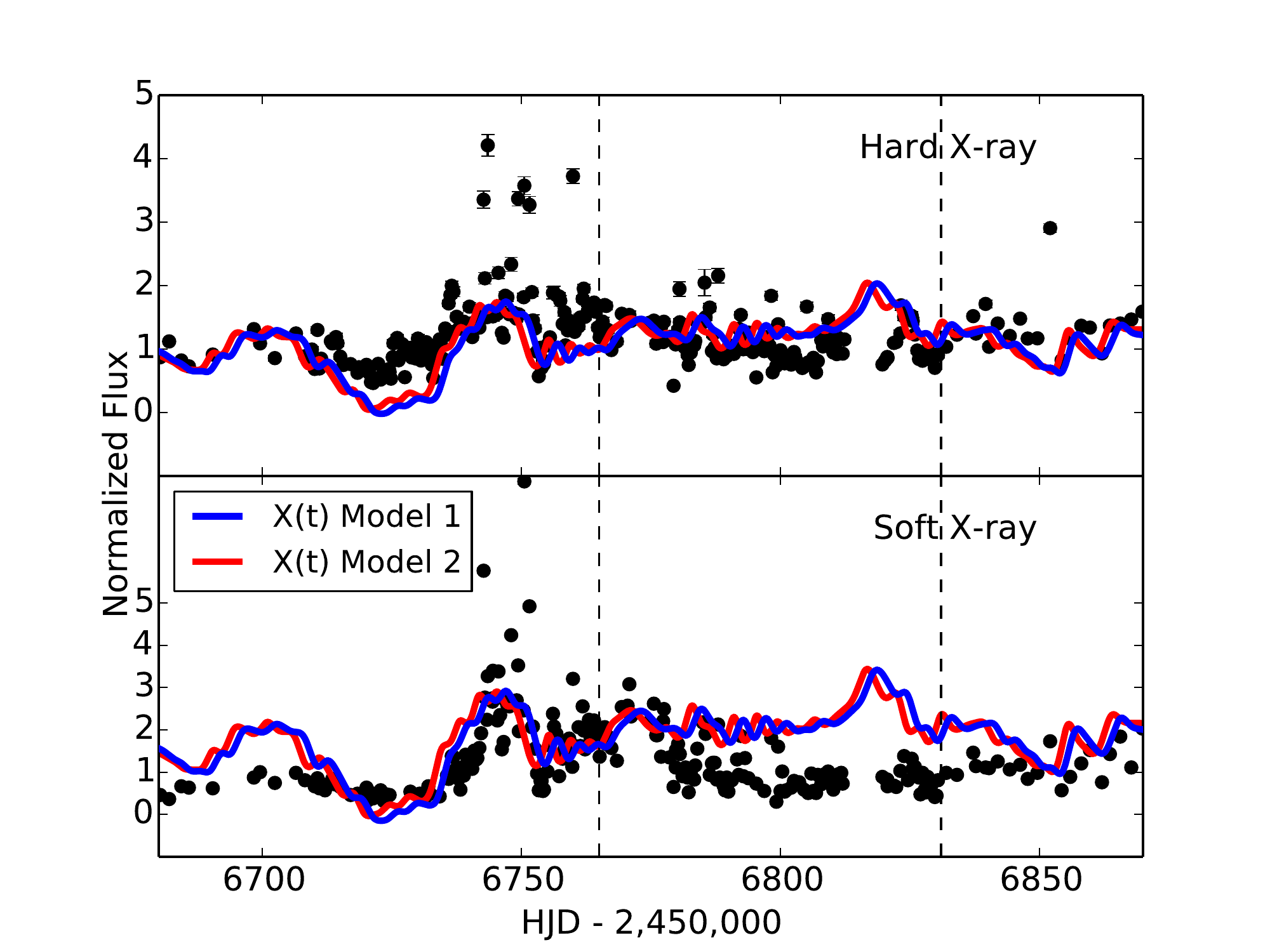}
\caption{Comparison of \texttt{CREAM}'s inferred driving light curve and the \swift hard (upper) and soft (lower) X-ray light curves. The ordinate scale is normalized to the mean of the hard and soft X-ray light curves, respectively, for the upper and lower panels. Blue and red lines show the driving light curves inferred by Models 1 and 2, respectively (Table \ref{tab_CREAMres}). The model light curves are shifted and scaled to match the mean and RMS of the X-ray data. The dashed lines enclose the BLR anomaly (Paper IV).}
\label{fig_CREAMxrcomp}
\end{figure*}

Some studies of AGN variability have found that X-ray light curves lead UV and optical light curves \citep{sh14,mc16}, making X-rays a candidate for the driving light curve. Figure \ref{fig_CREAMxrcomp} compares the hard and soft X-ray light curves (Paper II) to \texttt{CREAM}'s inferred driving light curve (Models 1 and 2). \texttt{CREAM}'s driving light curve is dimensionless and normalized to $\langle X \left( t \right) \rangle = 0$ and $\langle X^2 \left( t \right) \rangle = 1$. To compare it with the \swift hard and soft X-ray light curves, we shift and scale the \texttt{CREAM} light curve to match the mean and root-mean square (RMS) of the hard and soft X-ray light curves in turn (Figure \ref{fig_CREAMxrcomp}). The correlation coefficients $r_c$ of 0.35 and 0.38 for the hard and soft X-ray light curves (respectively) indicate a weak positive correlation between the \texttt{CREAM} estimate of the driving light curve and the X-ray light curves. We note that excluding the period of anomalous broad-line region (BLR) variability (Paper IV) does not significantly improve the level of correlation. 

These findings support the conclusions in Papers II and III that the observed X-rays alone cannot drive NGC 5548's variability during this campaign. We also note that the driving light curves inferred by Models 1 and 2 exhibit similar time structure with a slight delay. This offset arises since Model 2 ($\alpha$ varied as a free parameter) prefers larger overall mean lags than Model 1. The large lags are enabled by the high value of $T_1$ inferred in Model 2 relative to Model 1.The resulting mean lags from Model 2 agree more closely with the CCF values (Paper III) than do those from Model 1.

One problem may be that the \swift data only extend up to energies of $\sim10$\,keV, while the full spectral energy distribution (SED) peaks at $\sim100$\,keV \citep{ka14}. This may mean that the \swift observations are not a suitable proxy for the driving light curve. It is also observed by \citet{ga16} that, even smoothed by a response function, the hard X-ray light curve produces too much high-frequency variability to generate the UV and optical continuum light curves in NGC 5548. \citet{ga16} invoke an X-ray reprocessing region between the X-rays and accretion disk that may also be responsible for the poor correlations between the \texttt{CREAM} and X-ray light curves. We also speculate here that the accretion disk itself may well see the X-ray emission, but the observer does not. The poor correlation observed between the X-ray and UV light curves may simply be the effect of some absorbing medium between the X-ray-emitting corona and the observer that shields the X-ray emission from view but leaves it visible to the disk. Despite these observations, we note that models with the X-rays as the driving light curve can also work well \citep{tr16}. NGC 5548 is therefore clearly different in that respect.

\FloatBarrier

\pagebreak
\section{Accretion-Disk Spectrum}
\label{sec_fluxflux}

\begin{figure}
\centering
\includegraphics[scale=0.74,trim=0.5cm 0cm 0cm 0cm]{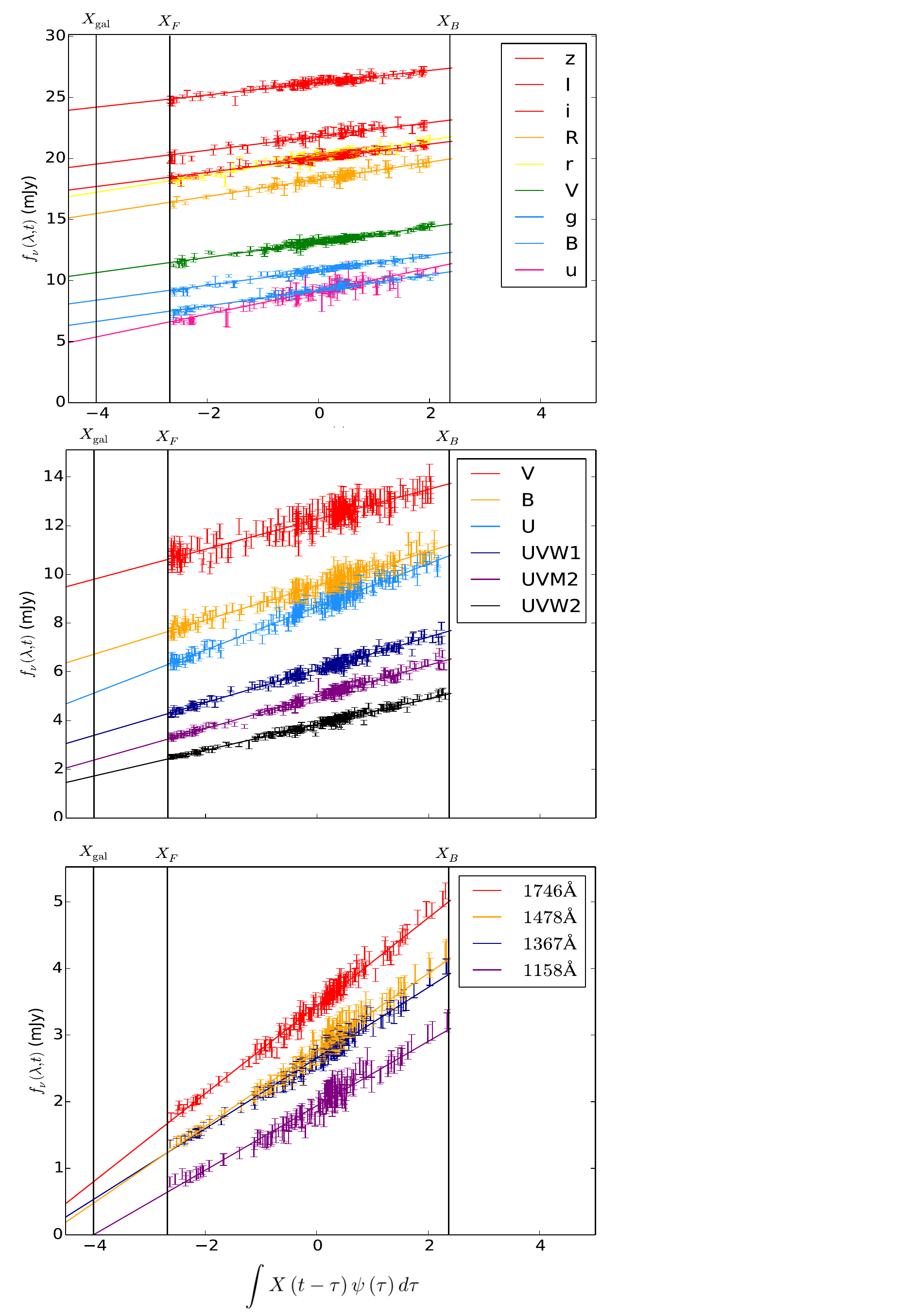}
\caption{The model response light curves as a function of the model driving light curve for the \texttt{CREAM} fits to the STORM light curves. The vertical lines label the driving light-curve values used to evaluate the galaxy, faint-state, and bright-state spectra ($X_\mathrm{gal}$, $X_F$, $X_B$). The upper, middle, and lower plots show the \hst, \swift, and ground-based light curves, respectively.}
\label{figfluxflux}
\end{figure}

The response light curves are compared against the driver $X(t)$ in Figure \ref{figfluxflux}. We note crucially that the accretion-disk variations from the faint state through to the bright state are linear. The lack of curvature validates the use of the linearized echo model (Equation \ref{eqfnu}) to fit the light curves in Figures \ref{figcreamfit1} to \ref{figcreamfit3}. Other features of interest in Figure \ref{figfluxflux} are the apparent steepness of the fit to both the ground-based \textit{u} and \swift \textit{U} light curves relative to the other wavelengths. This behavior could in part be due to Balmer continuum emission from the BLR (\citealt{ko01}, Paper III). A similar steepening in the \textit{r}-band light curve may be attributable to $\mathrm{H\alpha}$ BLR contamination (Paper III).

\begin{figure}
\centering
\includegraphics[scale=0.48,angle=0,trim=1cm 0cm 0cm 0cm]{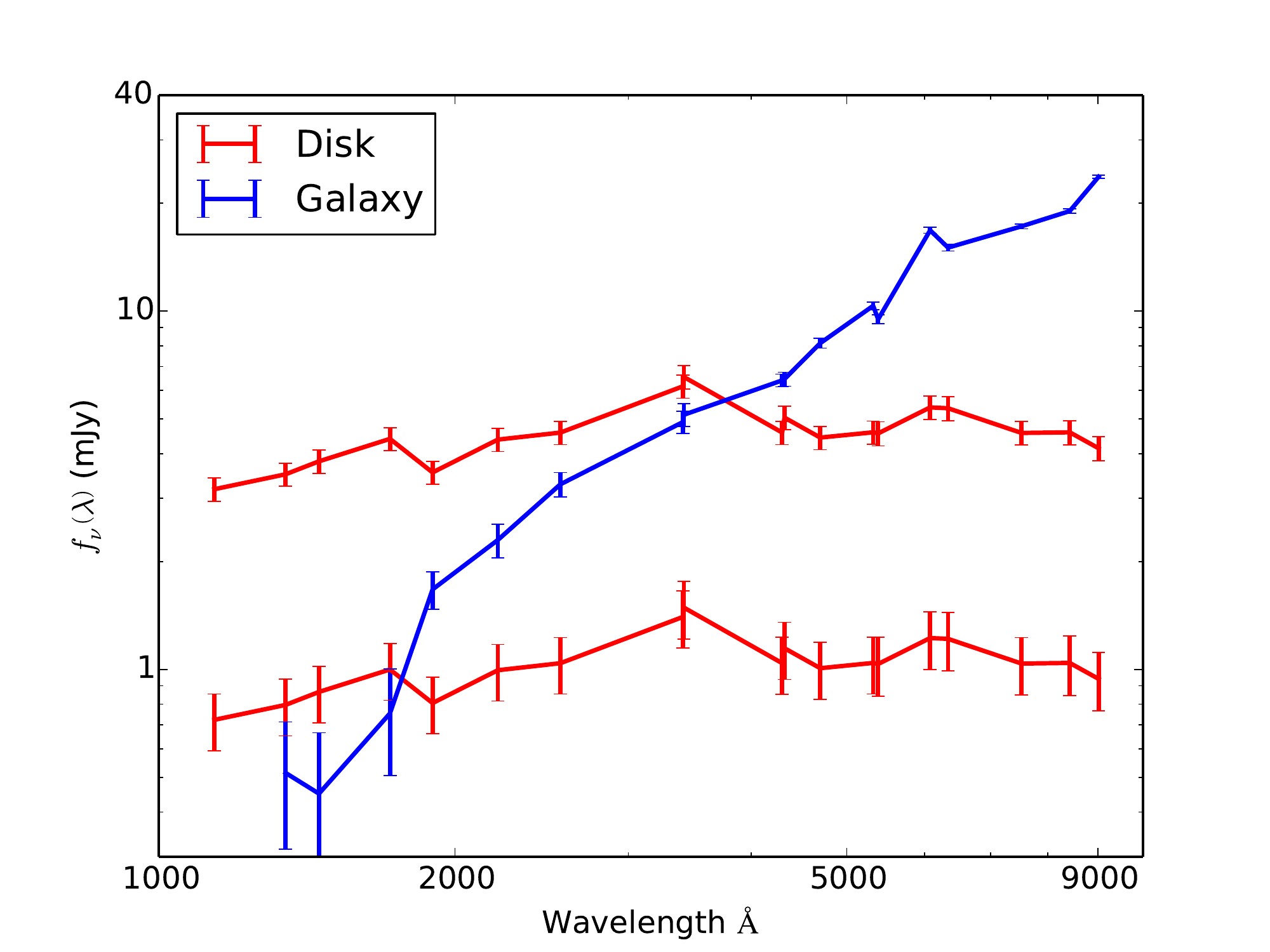}
\caption{The minimum and maximum (red error bars) disk spectra obtained following Equations \ref{eqfnufaint} and \ref{eqfnubright}. The host-galaxy spectrum is shown in blue following Equation \ref{eqfnugal}.} 
\label{fig_minmaxgal}
\end{figure}

We now use the linear relation between the driving and echo light-curve fluxes to estimate the accretion-disk spectrum using a technique similar to the flux vector gradient (FVG) method \citep{ca07,ha11}. The host-galaxy flux at the shortest wavelength is taken as the point where the linear fit crosses zero at the shortest wavelength (see the lower panel of Figure \ref{figfluxflux}) at $X \left( t \right) = X_\mathrm{gal}$. The host-galaxy contribution at the larger wavelengths is given by the linear trend lines evaluated at $X_\mathrm{gal}$. The faintest and brightest points of the light curves $f_\nu \left( \lambda, t \right)$ yield corresponding faint and bright states for the driving light curve $X_F$ and $X_B$ with which to evaluate a faint- and bright-state accretion-disk spectrum. The galaxy, faint-state disk $f_{\nu}^F$, and bright-state disk $f_{\nu}^B$ spectra are shown in Figure \ref{fig_minmaxgal} and are obtained using

\begin{equation}
\label{eqfnugal}
f_\mathrm{gal} \left( \lambda \right) = \bar{F}_{\nu} \left( \lambda \right)  + X_\mathrm{gal} \Delta F_\nu \left( \lambda \right),
\end{equation}

\begin{equation}
\label{eqfnufaint}
f_\nu^F \left( \lambda \right) = \bar{F}_{\nu} \left( \lambda \right)  + X_\mathrm{F} \Delta F_\nu \left( \lambda \right) - f_\mathrm{gal} \left( \lambda \right),
\end{equation}

\noindent and

\begin{equation}
\label{eqfnubright}
f_\nu^B \left( \lambda \right) = \bar{F}_{\nu} \left( \lambda \right)  + X_\mathrm{B} \Delta F_\nu \left( \lambda \right) - f_\mathrm{gal} \left( \lambda \right).
\end{equation}

\begin{figure}
\centering
\includegraphics[scale=0.48,angle=0,trim=1cm 0cm 0cm 0cm]{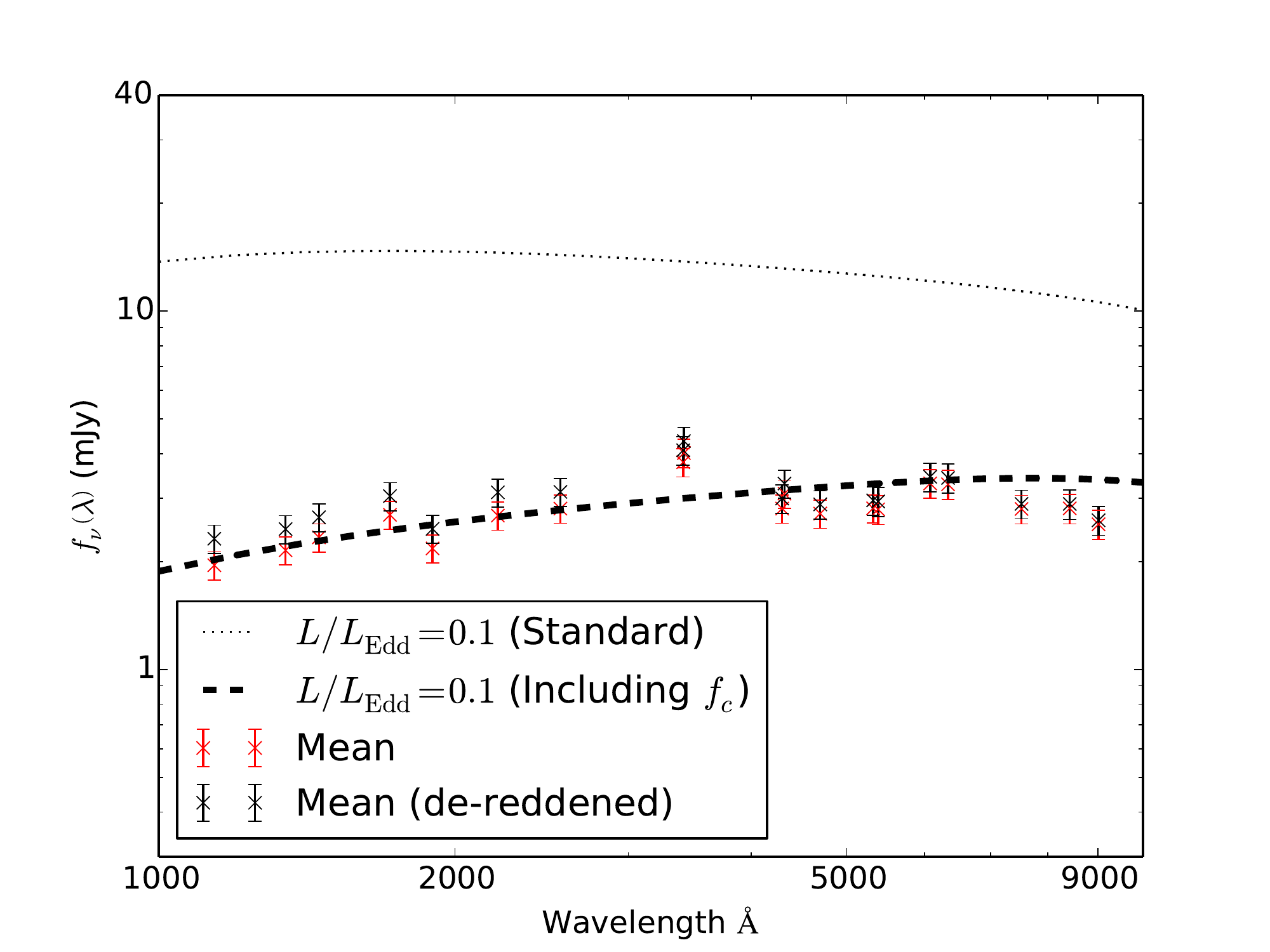}
\caption{The mean reddened (red points) and dereddened (black points) accretion-disk spectrum from Equations \ref{eqfnufaint} and \ref{eqfnubright}. The dotted line shows the spectrum for a standard blackbody disk with $L/L_\mathrm{Edd} = 0.1$. The thick dashed line shows the best-fit model for a partially covered accretion disk (Equation \ref{eq_covfrac}) fitted simultaneously to the mean spectra and the CCF lags (Figure \ref{fig_lagvswavelength}).} 
\label{fig_covfrac}
\end{figure}

\noindent Figure \ref{fig_covfrac} displays the mean accretion disk spectrum obtained by averaging the flux from Equations \ref{eqfnufaint} and \ref{eqfnubright}. For reference, we overlay a blackbody spectrum evaluated for a face-on disk, with $M_\mathrm{BH} = 10^{7.51}\,M_\odot$ \citep{pa14}, and adopt the Eddington ratio $L/L_\mathrm{Edd} = 0.1$ (Paper III). We correct for Milky Way extinction using \citet{se79} with an $E(B-V)$ parameter of 0.02\,mag \citep{ca07}.

Figures \ref{fig_minmaxgal} and \ref{fig_covfrac} show evidence that the minimum and maximum disk spectra turn down at short wavelengths. This could be evidence of the short-wavelength Wien cutoff predicted by blackbody models. The standard $L/L_\mathrm{Edd} = 0.1$ spectrum (dashed line in Figure \ref{fig_covfrac}) also displays this turndown but more slowly, and toward shorter wavelengths than the mean disk spectrum. Figure \ref{fig_covfrac} also shows that the accretion-disk spectrum appears in general too faint to be explained by a standard blackbody-emitting accretion disk with $L/L_\mathrm{Edd}=0.1$. This difference is the same as the ``flux size'' problem found in microlensing estimates of accretion-disk sizes \citep{mo10}.

We end this section by tentatively proposing a mechanism to account for both the large \texttt{CREAM} and CCF lags from Paper III, and the apparent faint accretion-disk spectrum. We suggest a modification to the standard blackbody-emitting disk. For a standard disk, the flux is simply the summation of blackbody curves of discrete annuli that increase in radius. We suggest that each annulus is only partially covered with blackbody-emitting material with a covering fraction $f_c \left( r  \right)$. We investigate a power-law covering fraction

\begin{equation}
\label{eq_covfrac}
f_c \left( r \right) = f_1 \left( \frac{r}{r_1} \right)^\gamma.
\end{equation}

\noindent The two parameters are the covering fraction $f_1$ at $r_1=1$ light day, and a slope parameter $\gamma$ that governs how quickly the disk transitions from fully blackbody-emitting annuli ($f_c=1$) to annuli with little or no blackbody emission ($f_c=0$). We incorporate the covering fraction into our calculation of the transfer function (Equation \ref{eqpsitaulam_A}) and then fit the mean lags at each of the STORM light-curve wavelengths. Using a simple grid search, simultaneously minimizing $\chi^2$ of the CCF lags (Paper III and Figure \ref{fig_lagvswavelength}) and the mean disk spectra (Figure \ref{fig_covfrac}), we optimize the covering fraction parameters $r_1$ and $\gamma$ with mean lags and uncertainties from the CCF analyses of Paper III. The best-fit values and uncertainties are $r_1=0.34 \pm 0.01$ and $\gamma = 0.33 \pm 0.03$. The resulting lag and disk spectra are shown by the dashed lines in Figures \ref{fig_lagvswavelength} and \ref{fig_covfrac}. Figure \ref{fig_lagvswavelength} demonstrates that incorporating a partially covered accretion disk can go at least some way to reconciling the difference in the observed and expected time lags. We see also from Figure \ref{fig_covfrac} that the accretion-disk spectrum is too faint to be explained by a standard blackbody accretion disk with $L/L_\mathrm{Edd} = 0.1$, and that a partial blackbody disk model can again go some way to account for this. 

Despite these results, we acknowledge that other authors find somewhat lower Eddington ratios for this object. For example, \citet{br12} find $L/L_\mathrm{Edd} = 0.08$ and \citet{ho14} argue for a value as low as $L/L_\mathrm{Edd} = 0.01$. It may therefore be that partial covering is not required to dim a standard disk model to match the mean disk spectrum, and is needed only to reconcile the long lags with the standard disk model. We also note that Paper III performs more sophisticated spectral decomposition techniques to estimate the host-galaxy contribution and finds a fainter galaxy (and thus brighter accretion disk) SED than that presented here. This further reduces the need for a covered disk model to the point where such a model may only be needed to explain the time lags rather than also being required to dim the flux spectrum. Further investigation of the covering fraction is well beyond the scope of this work but may form the basis of future studies.

The large lag spectrum (Figure \ref{fig_lagvswavelength}) may also support the findings of \citet{ga16}, whereby the disk does not respond to a single point-source driving light curve, but rather to a diffuse region that intercepts X-ray photons and re-emits these onto the accretion disk. This hypothesis is further supported by the poor correlation between the \texttt{CREAM}-inferred driving light curve and the \swift X-ray light curves.

\FloatBarrier

\section{Discussion}

We fit the 19 AGN STORM continuum light curves of \ngc with our \textbf{C}ontinuum \textbf{RE}procesing \textbf{A}GN \textbf{M}CMC code \texttt{CREAM}. \texttt{CREAM} assumes that accretion-disk time lags arise due to thermal reprocessing of irradiating photons from a time-varying point source (lamppost). \texttt{CREAM} models the lamppost light curve as a sum of Fourier sine and cosine terms with amplitudes constrained by a random-walk prior. The code requires one or more input continuum light curves and fits a response function for each wavelength parametrized by inclination, temperature $T_1$ at 1 light day, and slope of the temperature-radius profile. We first fit the light curves assuming a steady-state disk ($T \propto r^{-\alpha}$, where $\alpha \equiv 3/4$) and find best-fit values of $T_1 = \left( 22.2 \pm 0.7 \right) \times 10^3$\,K and $i = 54^\circ \pm 6^\circ$. If we relax the thin-disk temperature-radius law, the inferred disk parameters are $T_1 = \left( 47 \pm 5 \right) \times 10^3$\,K, $i = 37^\circ \pm 10^\circ$, and $\alpha = 0.99 \pm 0.03$. Such a steeper fall-off of temperature with radius than expected from a steady-state disk leads to a mean lag spectrum broadly in agreement with Paper III.

In general, the lamppost model seems to fit the continuum light curves from the 2014 STORM monitoring campaign relatively well. While there is a period of anomalous variability in the broad lines (Paper IV), this does not appear to be repeated in the continuum light curves. We also find that flux variations across all the continuum light curves from faint to bright states are linear, again consistent with a disk-reprocessing model. On the other hand, our analysis infers properties of the disk that are not consistent with X-ray reprocessing by a standard blackbody accretion disk. We first note from Figure \ref{fig_CREAMxrcomp} that the X-rays, in this observing campaign, appear to be a poor proxy for the driving light curve, as they do not correlate well with the continuum driving light curve inferred from \texttt{CREAM}. We also find large mean delays in the light-curve transfer functions across the \swift\ and ground-based data, in agreement with Paper III, and note the presence of a fainter accretion-disk spectrum than that which would be predicted assuming a standard blackbody accretion disk emitting at $L/L_\mathrm{Edd} = 0.1$.

We suggest here that the large observed mean lags from both \texttt{CREAM} and CCF techniques  may be the result of an accretion disk that is only partially covered with blackbody-emitting material with a covering fraction that increases with radius. While this model can explain the large time lags, it also requires an accretion disk that is much dimmer than observed here and in Paper III. A \citet{ga16} model may be the answer here, in which double reprocessing of X-rays occur. This will also give rise to the large time lags we see in Figure \ref{fig_lagvswavelength} relative to the driving light curve, since photons require multiple light-travel paths before intercepting the accretion disk. Another interpretation is that the inner disk is tilted relative to the outer disk, aligning itself with the spin of the black hole inward of some radius \citep{ne15}. The implication of this interpretation is that the driving light curve is actually just the emission from the inner tilted disk. This again introduces the concept of double reprocessing in that the X-rays initially reverberate from the inner accretion disk, which then irradiates the outer disk.

We again emphasize that the origin of the driving light curve remains ambiguous and requires further investigation of the correlations and lags of X-ray, UV, and optical continuum light curves. Be it an inner cylindrical X-ray reprocessing medium \citep{ga16}, an inner tilted accretion disk model \citep{ne15}, or something not yet conceived, it seems that our understanding of the driving light-curve mechanism will improve considerably in the near future.

\section{Acknowledgments}

Support for {\it HST} program number GO-13330 was provided by NASA through a grant from the Space Telescope Science Institute, which is operated by the Association of Universities for Research in Astronomy, Inc.,  under NASA contract NAS5-26555. D.A.S. and K.D.H. acknowledge support from the UK Science and Technology Facilities Council through grant ST/K502339/1 and ST/J001651/1. M.M.F., G.D.R., B.M.P., C.J.G., and R.W.P. are grateful for the support of the National Science Foundation (NSF) through grant AST-1008882 to The Ohio State University. A.J.B. and L.P. have been supported by NSF grant AST-1412693. A.V.F. and W.Z. are grateful for financial assistance from NSF grant AST-1211916, the TABASGO Foundation, and the Christopher R. Redlich Fund.   M.C. Bentz gratefully acknowledges support through NSF CAREER grant AST-1253702 to Georgia State University. M.C. Bottorff acknowledges HHMI for support through an undergraduate science education grant to Southwestern University.  K.D.D. is supported by an NSF Fellowship awarded under grant AST-1302093.  R.E. gratefully acknowledges support from NASA under awards NNX13AC26G, NNX13AC63G, and NNX13AE99G. J.M.G. gratefully acknowledges support from NASA under award NNX15AH49GHi Jonathan. P.B.H. is supported by NSERC.  M.I. acknowledges support from the Creative Initiative program, No.  20080060544, of the National Research Foundation of Korea (NRFK) funded by the Korean government (MSIP).
M.D.J. acknowledges NSF grant AST-0618209. SRON is financially supported by NWO, the Netherlands Organization for Scientific Research. B.C.K. is partially supported by the UC Center for Galaxy Evolution. C.S.K. acknowledges the support of NSF grant AST-1515876. D.C.L. acknowledges support from NSF grants AST-1009571 and AST-1210311. P.L. acknowledges support from Fondecyt grant 1120328. A.P. acknowledges support from an NSF graduate fellowship and a UCSB Dean’s Fellowship. J.S.S. acknowledges CNPq, National Council for Scientific and Technological Development (Brazil) for partial support and The Ohio State University for warm hospitality. T.T. has been supported by NSF grant AST-1412315. T.T. and B.C.K. acknowledge support from the Packard Foundation in the form of a Packard Research Fellowship to T.T.; also, T.T. thanks the American Academy in Rome and the Observatory of Monteporzio Catone for kind hospitality. The Dark Cosmology Centre is funded by the Danish National Research Foundation.  M.V. gratefully acknowledges support from the Danish Council for Independent Research via grant no.  DFF  4002-00275.  J.-H.W. acknowledges support by the National Research Foundation of Korea (NRF) grant funded by the Korean government (No. 2010-0027910). This research has made use of the NASA/IPAC Extragalactic Database (NED), which is operated by the Jet Propulsion Laboratory, California Institute of Technology, under contract with NASA. The authors acknowledge the support of the referee for their helpful input during the review process. 

\label{section:disco}

\appendix
\label{secappendix}

\section{Response Function}

The disk is modelled as a blackbody with local temperature $T$ determined by the viscous heating and lamppost irradiation. Variable lamppost irradiation is realized by the disk surface after a delay $\tau$ that is a function of inclination and radius. The flux is the integral of the Planck function $B_{\nu}(T)$ over surface elements with solid angle $d\Omega$,

\begin{equation}
\label{fnu}
F_\nu \left(\lambda, t \right) = \int B_{\nu} ( \lambda, T (t - \tau) ) d\Omega.
\end{equation}
$T(t-\tau)$ is the disk temperature at the appropriate look-back time $\tau$. We then must sum this over all points in the disk, each with its own time delay. If $X \left( t \right)$ is the driving light curve, the resulting convolution integral takes the form 

\begin{equation}
F_{\nu}(\lambda, t) = \bar{F_\nu} (\lambda) + \Delta F_\nu \left( \lambda \right) \int_{0}^{\infty} \psi ({\tau} | \lambda) X \left( t - \tau \right) d\tau.
\end{equation}

Differentiating with respect to the driving light curve, $X \left( t - \tau \right)$ gives
\begin{equation}
\label{halfwaytopsi}
 \frac {\partial F_\nu \left(\lambda, t \right)}{\partial X \left( t - \tau \right)} = \int_{0}^{\infty} d \tau' \psi( \tau' |\lambda ) \delta (\tau - \tau' ) = \int d \Omega \frac {\partial B_{\nu}}{\partial T} \frac{\partial T}{\partial X} \delta (\tau - \tau' (r,\theta, i) ),
 \end{equation} 
 
\noindent where the delta function considers only disk surface elements, $\tau' \left( r, \theta \right)$, with a common delay $\tau$. The result is  
 
 \begin{equation}
 \label{eqpsitaulam_A}
 \psi_\nu (\tau | \lambda ) = \int d \Omega \frac{\partial B_{\nu} (T, \lambda ) }{\partial T} \frac{\partial T}{\partial L_x } \frac{\partial L_x }{\partial F_x } \delta (\tau - \tau' (r, \theta, i ) ). 
 \end{equation}

\texttt{CREAM} evaluates this integral numerically as a sum over a grid that is logarithmic in radius and uniform in azimuth,

\begin{equation}
\begin{split}
\label{eqpsitaulam_A}
\psi_\nu (\tau | \lambda ) = \sum_\mathrm{ir} \sum_\mathrm{ia} \Delta \Omega \frac{\partial B_{\nu} (T \left( r_\mathrm{ir}, \theta_\mathrm{ia} \right), \lambda ) }{\partial T} \frac{\partial T}{\partial L_x } \frac{\partial L_x }{\partial F_x } f_c\left( r_\mathrm{ir} \right) \delta (\tau - \tau' (r_\mathrm{ir}, \theta_\mathrm{ia}, i ) ),
\end{split}
\end{equation}
 
\noindent where $\Delta \Omega = r_\mathrm{ir} \Delta r_\mathrm{ir} \Delta \theta / D_L^2$ is the solid-angle element, $D_L$ is the luminosity distance, and $F_x$ and $L_x$ are the driving light curve $X \left( t \right)$ expressed as a flux and luminosity (respectively). $f_c$ is the covering fraction introduced in Section \ref{sec_fluxflux}. Since we are already normalising $\psitau$ using Equation \ref{eq_psinorm}, the $\partial L_x / \partial F_x$ term is present only as a conceptual aid, introducing a constant factor of $4 \pi$ to the response function that does not affect the final result. In this work, the response function is evaluated up to a maximum delay of 50 days.

\FloatBarrier

\begin{table} 
\center 
\caption{Error bar expansion factor $f$ inferred by \texttt{CREAM} for the \textit{HST} and \swift light curves.}
\begin{tabular}{cccccc} 
\hline 
Telescope & Filter & $f$ & $\langle \sigma \rangle \times f $ & $N$ & $\chi^2/(Nf^2)$ \\ 
 & & & mJy & & \\\hline
\hline 
HST & $1158\,\mathrm{\AA}$ & $2.41 \pm 0.15$ & $0.11 \pm 0.01$ & $171$ & $1.15$ \\ 
HST & $1367\,\mathrm{\AA}$ & $1.92 \pm 0.13$ & $0.09 \pm 0.01$ & $171$ & $0.90$ \\ 
HST & $1478\,\mathrm{\AA}$ & $2.37 \pm 0.15$ & $0.10 \pm 0.01$ & $171$ & $0.87$ \\ 
HST & $1746\,\mathrm{\AA}$ & $1.18 \pm 0.08$ & $0.09 \pm 0.01$ & $171$ & $0.98$ \\ 
Swift & UVW2 & $1.08 \pm 0.05$ & $0.09 \pm 0.00$ & $284$ & $0.96$ \\ 
Swift & UVM2 & $0.89 \pm 0.04$ & $0.12 \pm 0.01$ & $256$ & $1.00$ \\ 
Swift & UVW1 & $0.85 \pm 0.04$ & $0.13 \pm 0.01$ & $270$ & $0.82$ \\ 
Swift & \textit{U} & $0.82 \pm 0.04$ & $0.20 \pm 0.01$ & $270$ & $0.99$ \\ 
Swift & \textit{B} & $0.85 \pm 0.04$ & $0.22 \pm 0.01$ & $271$ & $1.02$ \\ 
Swift & \textit{V} & $0.78 \pm 0.04$ & $0.32 \pm 0.01$ & $260$ & $1.21$ \\ 
\hline 
\end{tabular}
\footnote{$f$ is the error bar scale factor, $\langle \sigma \rangle$ is the mean error bar at each wavelength, and $N$ is the number of data points for each telescope-filter combination.} 
\label{tabsigexpand} 
\end{table}

\begin{table} 
\center 
\caption{As in Table \ref{tabsigexpand}, but for the ground-based telescopes.}
\begin{tabular}{cccccc} 
\hline 
Telescope \footnote{ Acronyms --- CAO: Crimean Astrophysical Observatory; LOKAIT: Lick Observatory Katzman Automatic Imaging Telescope; LT: Liverpool Telescope; LCO/SSO: Siding Spring Observatory; LCO/McD: McDonald; WMO: West Mountain Obsrvatory; RCT: Robotically Controlled Telescope; HLC: Hard Labour Creek Observatory; MLOAO: Mt. Lemmon Optical Astronomy Observatory; MLO: Mt. Laguna Observatory; FO: Fountainwood Observatory; LCO/SAAO: South African Astronomical Observatory} & Filter & $f$ & $\langle \sigma \rangle \times f $ & $N$ & $\chi^2/(Nf^2)$ \\ 
 & & & (mJy) & & \\\hline
\hline 
LT & \textit{u} & $8.58 \pm 0.65$ & $0.23 \pm 0.02$ & $103$ & $1.14$ \\ 
LCO/McD & \textit{u} & $0.90 \pm 0.12$ & $0.54 \pm 0.07$ & $35$ & $0.83$ \\ 
LCO/SAAO a & \textit{u} & $3.37 \pm 1.13$ & $0.65 \pm 0.23$ & $7$ & $1.80$ \\ 
Wise & \textit{B} & $2.50 \pm 0.28$ & $0.09 \pm 0.01$ & $58$ & $0.79$ \\ 
LCO/SAAO a & \textit{B} & $1.38 \pm 2.01$ & $0.25 \pm 0.37$ & $2$ & $1.06$ \\ 
LCO/SAAO b & \textit{B} & $0.78 \pm 0.34$ & $0.17 \pm 0.07$ & $5$ & $1.54$ \\ 
LCO/SAAO c & \textit{B} & $0.88 \pm 0.46$ & $0.18 \pm 0.10$ & $4$ & $1.40$ \\ 
CAO & \textit{B} & $1.31 \pm 0.15$ & $0.21 \pm 0.03$ & $44$ & $1.17$ \\ 
LCO/McD & \textit{B} & $1.08 \pm 0.48$ & $0.09 \pm 0.04$ & $5$ & $0.17$ \\ 
RCT & \textit{B} & $1.57 \pm 0.23$ & $0.10 \pm 0.02$ & $30$ & $0.78$ \\ 
WMO & \textit{B} & $2.51 \pm 1.97$ & $0.25 \pm 0.21$ & $3$ & $0.14$ \\ 
LT & \textit{g} & $6.02 \pm 0.45$ & $0.13 \pm 0.01$ & $104$ & $0.93$ \\ 
LCO/SSO a & \textit{g} & $0.64 \pm 0.18$ & $0.20 \pm 0.06$ & $9$ & $1.56$ \\ 
LCO/SSO b & \textit{g} & $0.76 \pm 0.23$ & $0.25 \pm 0.08$ & $8$ & $0.91$ \\ 
LCO/SAAO a & \textit{g} & $1.00 \pm 0.32$ & $0.10 \pm 0.03$ & $8$ & $0.75$ \\ 
LCO/McD & \textit{g} & $1.14 \pm 0.13$ & $0.19 \pm 0.02$ & $43$ & $0.80$ \\ 
Wise & \textit{V} & $2.44 \pm 0.24$ & $0.09 \pm 0.01$ & $76$ & $0.88$ \\ 
FO & \textit{V} & $2.33 \pm 0.22$ & $0.15 \pm 0.01$ & $66$ & $0.73$ \\ 
LOKAIT & \textit{V} & $0.70 \pm 0.07$ & $0.20 \pm 0.02$ & $61$ & $0.87$ \\ 
McD & \textit{V} & $1.90 \pm 0.22$ & $0.14 \pm 0.02$ & $46$ & $1.12$ \\ 
Maidanak & \textit{V} & $1.77 \pm 0.25$ & $0.16 \pm 0.02$ & $30$ & $0.89$ \\ 
WMO & \textit{V} & $1.26 \pm 0.18$ & $0.12 \pm 0.02$ & $31$ & $0.80$ \\ 
RCT & \textit{V} & $2.89 \pm 0.42$ & $0.13 \pm 0.02$ & $31$ & $0.58$ \\ 
HLCO & \textit{V} & $1.45 \pm 0.26$ & $0.09 \pm 0.02$ & $20$ & $0.89$ \\ 
MLO & \textit{V} & $1.78 \pm 0.68$ & $0.25 \pm 0.10$ & $6$ & $0.59$ \\ 
MLOAO & \textit{V} & $2.36 \pm 0.45$ & $0.16 \pm 0.03$ & $17$ & $1.12$ \\ 
LCO/SAAO b & \textit{V} & $1.99 \pm 0.52$ & $0.17 \pm 0.05$ & $10$ & $0.93$ \\ 
CTIO 1 & \textit{V} & $0.61 \pm 0.21$ & $0.16 \pm 0.05$ & $7$ & $1.31$ \\ 
CTIO 2 & \textit{V} & $0.43 \pm 0.10$ & $0.07 \pm 0.02$ & $13$ & $1.87$ \\ 
LCO/SAAO a & \textit{V} & $1.31 \pm 0.40$ & $0.12 \pm 0.04$ & $8$ & $0.96$ \\ 
LCO/SSO b & \textit{V} & $1.69 \pm 0.73$ & $0.18 \pm 0.08$ & $5$ & $0.57$ \\ 
LCO/SSO a & \textit{V} & $1.10 \pm 2.17$ & $0.10 \pm 0.19$ & $2$ & $0.87$ \\ 
LT & \textit{r} & $6.10 \pm 0.48$ & $0.18 \pm 0.02$ & $99$ & $1.12$ \\ 
LCO/SSO b & \textit{r} & $0.70 \pm 0.24$ & $0.37 \pm 0.13$ & $7$ & $1.17$ \\ 
LCO/SAAO a & \textit{r} & $0.39 \pm 0.12$ & $0.31 \pm 0.10$ & $8$ & $1.41$ \\ 
LCO/SAAO c & \textit{r} & $1.41 \pm 0.31$ & $0.22 \pm 0.05$ & $14$ & $1.47$ \\ 
LCO/McD & \textit{r} & $1.84 \pm 0.22$ & $0.27 \pm 0.03$ & $44$ & $1.07$ \\ 
Wise & \textit{R} & $4.36 \pm 0.45$ & $0.19 \pm 0.02$ & $60$ & $1.50$ \\ 
CAO & \textit{R} & $1.31 \pm 0.17$ & $0.36 \pm 0.05$ & $34$ & $1.02$ \\ 
Maidanak & \textit{R} & $1.27 \pm 0.70$ & $0.16 \pm 0.09$ & $4$ & $0.44$ \\ 
WMO & \textit{R} & $2.96 \pm 0.37$ & $0.20 \pm 0.03$ & $38$ & $0.88$ \\ 
LT & \textit{i} & $6.58 \pm 0.50$ & $0.17 \pm 0.01$ & $108$ & $0.94$ \\ 
LCO/SAAO b & \textit{i} & $1.09 \pm 0.61$ & $0.20 \pm 0.11$ & $4$ & $0.53$ \\ 
LCO/SAAO a & \textit{i} & $2.80 \pm 1.25$ & $0.28 \pm 0.14$ & $5$ & $0.31$ \\ 
LCO/SAAO c & \textit{i} & $2.55 \pm 0.52$ & $0.25 \pm 0.06$ & $16$ & $1.04$ \\ 
LCO/McD & \textit{i} & $1.49 \pm 0.18$ & $0.17 \pm 0.02$ & $45$ & $1.21$ \\ 
Wise & \textit{I} & $4.56 \pm 0.44$ & $0.23 \pm 0.03$ & $64$ & $0.99$ \\ 
CAO & \textit{I} & $1.16 \pm 0.15$ & $0.35 \pm 0.05$ & $34$ & $1.64$ \\ 
LT & \textit{z} & $5.26 \pm 0.40$ & $0.20 \pm 0.02$ & $108$ & $0.87$ \\ 
LCO/SAAO 1 & \textit{z} & $0.83 \pm 0.24$ & $0.23 \pm 0.07$ & $9$ & $0.58$ \\ 
LCO/SAAO 1 & \textit{z} & $0.22 \pm 0.08$ & $0.16 \pm 0.06$ & $7$ & $0.28$ \\ 
LCO/SAAO 3 & \textit{z} & $1.06 \pm 0.20$ & $0.25 \pm 0.05$ & $17$ & $0.93$ \\ 
LCO/McD & \textit{z} & $1.05 \pm 0.12$ & $0.21 \pm 0.03$ & $45$ & $1.26$ \\ 
\hline 
\end{tabular} 
\label{tabsigexpand_2} 
\end{table}

\begin{table} 
\center 
\caption{Mean and RMS spectra from \texttt{CREAM}.} 
\begin{tabular}{ccc} 
\hline 
 & $\bar{F}_\nu \left( \lambda \right)$ & $\Delta F_\nu \left( \lambda \right)$ \\ 
$\lambda$ & \multicolumn{2}{c}{\texttt{CREAM}} \\
(\AA) & \multicolumn{2}{c}{(mJy)}  \\
\hline 
\hline 
$1158$ & $1.82 \pm 0.15$ & $0.44 \pm 0.02$ \\ 
$1367$ & $2.53 \pm 0.15$ & $0.48 \pm 0.02$ \\ 
$1478$ & $2.64 \pm 0.18$ & $0.53 \pm 0.02$ \\ 
$1746$ & $3.28 \pm 0.21$ & $0.61 \pm 0.02$ \\ 
$1928$ & $3.71 \pm 0.15$ & $0.49 \pm 0.02$ \\ 
$2246$ & $4.81 \pm 0.21$ & $0.61 \pm 0.02$ \\ 
$2600$ & $5.91 \pm 0.21$ & $0.64 \pm 0.02$ \\ 
$3467$ & $8.44 \pm 0.30$ & $0.87 \pm 0.03$ \\ 
$3472$ & $9.35 \pm 0.24$ & $0.71 \pm 0.03$ \\ 
$4369$ & $9.10 \pm 0.21$ & $0.64 \pm 0.02$ \\ 
$4392$ & $8.89 \pm 0.30$ & $0.92 \pm 0.03$ \\ 
$4776$ & $10.79 \pm 0.21$ & $0.64 \pm 0.02$ \\ 
$5376$ & $12.08 \pm 0.21$ & $0.62 \pm 0.02$ \\ 
$5404$ & $12.97 \pm 0.21$ & $0.64 \pm 0.02$ \\ 
$6176$ & $19.88 \pm 0.24$ & $0.76 \pm 0.03$ \\ 
$6440$ & $18.09 \pm 0.27$ & $0.77 \pm 0.03$ \\ 
$7648$ & $19.85 \pm 0.21$ & $0.65 \pm 0.02$ \\ 
$8561$ & $21.63 \pm 0.24$ & $0.66 \pm 0.03$ \\ 
$9157$ & $26.05 \pm 0.21$ & $0.59 \pm 0.02$ \\ 
\hline 
\end{tabular} 
\label{tab_cream_fluxflux} 
\end{table}

\begin{table} 
\center 
\caption{Galaxy and disk spectra from the analysis in Section \ref{sec_fluxflux}.} 
\begin{tabular}{ccccc} 
\hline
$\lambda$ & $f_\nu^\mathrm{gal}$ & $f_\nu^\mathrm{F}$ & $f_\nu^\mathrm{B}$ & $f_\nu^B / f_\nu^F$ \\
(\AA) & (mJy) & (mJy) & (mJy) & \\
\hline 
\hline
$1158$ & $0$ & $0.77\pm0.02$ & $2.95\pm0.03$ & $3.84\pm0.03$ \\
$1367$ & $0.48\pm 0.03$ & $0.85\pm0.02$ & $3.27\pm0.03$ & $3.84\pm0.02$ \\
$1478$ & $0.41\pm 0.03$ & $0.93\pm0.02$ & $3.55\pm0.03$ & $3.84\pm0.02$ \\
$1746$ & $0.70\pm 0.03$ & $1.07\pm0.02$ & $4.12\pm0.03$ & $3.84\pm0.02$ \\
$1928$ & $1.64\pm 0.02$ & $0.81\pm0.01$ & $3.29\pm0.02$ & $4.08\pm0.01$ \\
$2246$ & $2.27\pm 0.02$ & $0.99\pm0.02$ & $4.04\pm0.02$ & $4.08\pm0.02$ \\
$2600$ & $3.27\pm 0.03$ & $1.03\pm0.02$ & $4.21\pm0.03$ & $4.08\pm0.02$ \\
$3467$ & $4.93\pm 0.04$ & $1.37\pm0.03$ & $5.60\pm0.04$ & $4.08\pm0.02$ \\
$3472$ & $5.14\pm 0.08$ & $1.46\pm0.07$ & $5.81\pm0.07$ & $3.97\pm0.03$ \\
$4369$ & $6.48\pm 0.04$ & $1.02\pm0.03$ & $4.08\pm0.03$ & $3.99\pm0.02$ \\
$4392$ & $6.53\pm 0.05$ & $1.10\pm0.03$ & $4.49\pm0.05$ & $4.08\pm0.03$ \\
$4776$ & $8.19\pm 0.04$ & $0.97\pm0.04$ & $3.86\pm0.04$ & $3.97\pm0.02$ \\
$5376$ & $10.32\pm0.04$ & $1.05\pm0.03$ & $4.20\pm0.03$ & $4.02\pm0.02$ \\
$5404$ & $9.66\pm 0.06$ & $0.95\pm0.05$ & $3.88\pm0.06$ & $4.08\pm0.04$ \\
$6176$ & $16.94\pm0.05$ & $1.22\pm0.04$ & $4.68\pm0.04$ & $3.84\pm0.02$ \\
$6440$ & $15.32\pm0.07$ & $1.19\pm0.06$ & $4.56\pm0.07$ & $3.84\pm0.04$ \\
$7648$ & $17.44\pm0.04$ & $0.95\pm0.04$ & $3.83\pm0.04$ & $4.02\pm0.03$ \\
$8561$ & $19.19\pm0.10$ & $0.97\pm0.07$ & $3.88\pm0.09$ & $3.99\pm0.06$ \\
$9157$ & $24.04\pm0.06$ & $0.85\pm0.05$ & $3.25\pm0.06$ & $3.84\pm0.04$ \\
\hline
\end{tabular}
\end{table}

\begin{table*}
\center
\caption{Response function mean lag vs. wavelength for \texttt{CREAM} models 11 and 2.}\label{tab_creamlag}
\begin{tabular}{ccccc}
\hline
& &  & \multicolumn{2}{c}{Model}\\
&&& 1 & 2 \\
Source & Bandpass & $\lambda_{\mathrm{pivot}}$ & \multicolumn{2}{c}{ } \\
& & $\left(\mathrm{\AA} \right)$ & \multicolumn{2}{c}{$\langle \tau \rangle$ (days)}\\
\hline
\hline 
{\it HST}    & & 1158   &      $0.47  \pm  0.01$   &   $1.94  \pm   0.17$      \\
{\it HST}    & & 1367   &      $0.57  \pm  0.01$   &   $2.07  \pm   0.15$      \\
{\it HST}    & & 1478   &      $0.63  \pm  0.01$   &   $2.15  \pm   0.14$      \\
{\it HST}    & & 1746   &      $0.77  \pm  0.01$   &   $2.32  \pm   0.13$      \\
{\it Swift}  & UVW2 & 1928   &     $0.87  \pm  0.02$   &   $2.44  \pm   0.12$      \\
{\it Swift}  & UVM2 & 2246   &     $1.06  \pm  0.02$   &   $2.64  \pm   0.11$      \\
{\it Swift}  & UVW1 & 2600   &     $1.28  \pm  0.02$   &   $2.87  \pm   0.10$      \\
{\it Swift}  & \textit{U} & 3467   &      $1.86  \pm  0.04$   &   $3.44  \pm   0.09$  \\
Ground & \textit{u} & 3472   &      $1.86  \pm  0.04$   &   $3.48  \pm   0.09$  \\
Ground & \textit{B} & 4369   &      $2.52  \pm  0.05$   &   $4.04  \pm   0.09$  \\
{\it Swift}  & \textit{B} & 4392   &      $2.54  \pm  0.05$   &   $4.05  \pm   0.09$  \\
Ground & \textit{g} & 4776   &      $2.84  \pm  0.06$   &   $4.31  \pm   0.08$  \\
{\it Swift}  & \textit{V} & 5376   &      $3.41  \pm  0.07$   &   $4.73  \pm   0.08$  \\
Ground & \textit{V} & 5404   &      $3.35 \pm  0.07$   &   $4.77  \pm   0.08$  \\
Ground & \textit{r} & 6176   &      $4.00  \pm  0.08$   &   $5.23  \pm   0.07$  \\
Ground & \textit{R} & 6440   &      $4.22  \pm  0.08$   &   $5.41  \pm   0.07$  \\
Ground & \textit{i} & 7648   &      $5.27  \pm  0.10$   &   $6.18  \pm   0.05$  \\
Ground & \textit{I} & 8561   &      $6.05  \pm  0.11$   &   $6.73  \pm   0.05$  \\
Ground & \textit{z} & 9157   &      $6.56  \pm  0.11$   &   $7.07  \pm   0.04$  \\

\hline
\end{tabular}
\end{table*}

\FloatBarrier

\bibliography{storm_mine} 

\begin{thebibliography}{}
\expandafter\ifx\csname natexlab\endcsname\relax\def\natexlab#1{#1}\fi

\bibitem[{{Alard} \& {Lupton}(1998)}]{al99}
{Alard}, C., \& {Lupton}, R.~H. 1998, apj, 503, 325

\bibitem[{{Blackburne} {et~al.}(2014){Blackburne}, {Kochanek}, {Chen}, {Dai},
  \& {Chartas}}]{bl14}
{Blackburne}, J.~A., {Kochanek}, C.~S., {Chen}, B., {Dai}, X., \& {Chartas}, G.
  2014, \apj, 789, 125

\bibitem[{{Blackburne} {et~al.}(2015){Blackburne}, {Kochanek}, {Chen}, {Dai},
  \& {Chartas}}]{bl15}
---. 2015, \apj, 798, 95

\bibitem[{{Blackburne} {et~al.}(2011){Blackburne}, {Pooley}, {Rappaport}, \&
  {Schechter}}]{bl11}
{Blackburne}, J.~A., {Pooley}, D., {Rappaport}, S., \& {Schechter}, P.~L. 2011,
  \apj, 729, 34

\bibitem[{{Blandford} \& {McKee}(1982)}]{bl82}
{Blandford}, R.~D., \& {McKee}, C.~F. 1982, apj, 255, 419

\bibitem[{{Brenneman} {et~al.}(2012){Brenneman}, {Elvis}, {Krongold}, {Liu}, \&
  {Mathur}}]{br12}
{Brenneman}, L.~W., {Elvis}, M., {Krongold}, Y., {Liu}, Y., \& {Mathur}, S.
  2012, \apj, 744, 13

\bibitem[{{Cackett} {et~al.}(2007){Cackett}, {Horne}, \& {Winkler}}]{ca07}
{Cackett}, E.~M., {Horne}, K., \& {Winkler}, H. 2007, mnras, 380, 669

\bibitem[{{Collier} {et~al.}(1998){Collier}, {Horne}, {Kaspi}, {Netzer},
  {Peterson}, {Wanders}, {Alexander}, {Bertram}, {Comastri}, {Gaskell},
  {Malkov}, {Maoz}, {Mignoli}, {Pogge}, {Pronik}, {Sergeev}, {Snedden},
  {Stirpe}, {Bochkarev}, {Burenkov}, {Shapovalova}, \& {Wagner}}]{co98}
{Collier}, S.~J., {Horne}, K., {Kaspi}, S., {et~al.} 1998, \apj, 500, 162

\bibitem[{{De Rosa} {et~al.}(2015){De Rosa}, {Peterson}, {Ely}, {Kriss},
  {Crenshaw}, {Horne}, {Korista}, {Netzer}, {Pogge}, {Arevalo}, {Barth},
  {Bentz}, {Brandt}, {Breeveld}, {Brewer}, {Dalla Bonta}, {De Lorenzo-Caceres},
  {Denney}, {Dietrich}, {Edelson}, {Evans}, {Fausnaugh}, {Gehrels}, {Gelbord},
  {Goad}, {Grier}, {Grupe}, {Hall}, {Kaastra}, {Kelly}, {Kennea}, {Kochanek},
  {Lira}, {Mathur}, {McHardy}, {Nousek}, {Pancoast}, {Papadakis}, {Pei},
  {Schimoia}, {Siegel}, {Starkey}, {Treu}, {Uttley}, {Vaughan}, {Vestergaard},
  {Villforth}, {Yan}, {Young}, \& {Zu}}]{ro15}
{De Rosa}, G., {Peterson}, B.~M., {Ely}, J., {et~al.} 2015, ArXiv e-prints,
  arXiv:1501.05954

\bibitem[{{Dexter} \& {Agol}(2011)}]{de11}
{Dexter}, J., \& {Agol}, E. 2011, \apjl, 727, L24

\bibitem[{{Edelson} {et~al.}(2015){Edelson}, {Gelbord}, {Horne}, {McHardy},
  {Peterson}, {Arevalo}, {Breeveld}, {DeRosa}, {Evans}, {Goad}, {Kriss},
  {Brandt}, {Gehrels}, {Grupe}, {Kennea}, {Kochanek}, {Nousek}, {Papadakis},
  {Siegel}, {Starkey}, {Uttley}, {Vaughan}, {Young}, {Barth}, {Bentz},
  {Brewer}, {Crenshaw}, {Dalla Bonta}, {De Lorenzo-Caceres}, {Denney},
  {Dietrich}, {Ely}, {Fausnaugh}, {Grier}, {Hall}, {Kaastra}, {Kelly},
  {Korista}, {Lira}, {Mathur}, {Netzer}, {Pancoast}, {Pei}, {Pogge},
  {Schimoia}, {Treu}, {Vestergaard}, {Villforth}, {Yan}, \& {Zu}}]{ed15}
{Edelson}, R., {Gelbord}, J.~M., {Horne}, K., {et~al.} 2015, ArXiv e-prints,
  arXiv:1501.05951

\bibitem[{{Fausnaugh} {et~al.}(2016){Fausnaugh}, {Denney}, {Barth}, {Bentz},
  {Bottorff}, {Carini}, {Croxall}, {De Rosa}, {Goad}, {Horne}, {Joner},
  {Kaspi}, {Kim}, {Klimanov}, {Kochanek}, {Leonard}, {Netzer}, {Peterson},
  {Schn{\"u}lle}, {Sergeev}, {Vestergaard}, {Zheng}, {Zu}, {Anderson},
  {Ar{\'e}valo}, {Bazhaw}, {Borman}, {Boroson}, {Brandt}, {Breeveld}, {Brewer},
  {Cackett}, {Crenshaw}, {Dalla Bont{\`a}}, {De Lorenzo-C{\'a}ceres},
  {Dietrich}, {Edelson}, {Efimova}, {Ely}, {Evans}, {Filippenko}, {Flatland},
  {Gehrels}, {Geier}, {Gelbord}, {Gonzalez}, {Gorjian}, {Grier}, {Grupe},
  {Hall}, {Hicks}, {Horenstein}, {Hutchison}, {Im}, {Jensen}, {Jones},
  {Kaastra}, {Kelly}, {Kennea}, {Kim}, {Korista}, {Kriss}, {Lee}, {Lira},
  {MacInnis}, {Manne-Nicholas}, {Mathur}, {McHardy}, {Montouri}, {Musso},
  {Nazarov}, {Norris}, {Nousek}, {Okhmat}, {Pancoast}, {Papadakis}, {Parks},
  {Pei}, {Pogge}, {Pott}, {Rafter}, {Rix}, {Saylor}, {Schimoia}, {Siegel},
  {Spencer}, {Starkey}, {Sung}, {Teems}, {Treu}, {Turner}, {Uttley},
  {Villforth}, {Weiss}, {Woo}, {Yan}, \& {Young}}]{fa16}
{Fausnaugh}, M.~M., {Denney}, K.~D., {Barth}, A.~J., {et~al.} 2016, \apj, 821,
  56

\bibitem[{Frank {et~al.}(2002)Frank, A.King, \& D.Raine}]{AP}
Frank, J., A.King, \& D.Raine. 2002, Accretion Power in Astrophysics
  (Cambridge, UK: Cambridge University Press)

\bibitem[{{Gardner} \& {Done}(2016)}]{ga16}
{Gardner}, E., \& {Done}, C. 2016, ArXiv e-prints, arXiv:1603.09564

\bibitem[{{Goad} {et~al.}(2016){Goad}, {Korista}, {De Rosa}, {Kriss},
  {Edelson}, {Barth}, {Ferland}, {Kochanek}, {Netzer}, {Peterson}, {Bentz},
  {Bisogni}, {Crenshaw}, {Denney}, {Ely}, {Fausnaugh}, {Grier}, {Gupta},
  {Horne}, {Kaastra}, {Pancoast}, {Pei}, {Pogge}, {Skielboe}, {Starkey},
  {Vestergaard}, {Zu}, {Anderson}, {Arevalo}, {Bazhaw}, {Borman}, {Boroson},
  {Bottorff}, {Brandt}, {Breeveld}, {Brewer}, {Cackett}, {Carini}, {Croxall},
  {Dalla Bonta}, {de Lorenzo-Caceres}, {Dietrich}, {Efimova}, {Evans},
  {Filippenko}, {Flatland}, {Gehrels}, {Geier}, {Gelbord}, {Gonzalez},
  {Gorjian}, {Grupe}, {Hall}, {Hicks}, {Horenstein}, {Hutchison}, {Im},
  {Jensen}, {Joner}, {Jones}, {Kaspi}, {Kelly}, {Kennea}, {Kim}, {Kim},
  {Klimanov}, {Larionov}, {Lee}, {Leonard}, {Lira}, {MacInnis},
  {Manne-Nicholas}, {Mathur}, {McHardy}, {Montouri}, {Musso}, {Nazarov},
  {Norris}, {Nousek}, {Okhmat}, {Papadakis}, {Parks}, {Pott}, {Rafter}, {Rix},
  {Saylor}, {Schimoia}, {Schnulle}, {Sergeev}, {Siegel}, {Spencer}, {Sung},
  {Teems}, {Treu}, {Turner}, {Uttley}, {Villforth}, {Weiss}, {Woo}, {Yan},
  {Young}, \& {Zheng}}]{go16}
{Goad}, M.~R., {Korista}, K.~T., {De Rosa}, G., {et~al.} 2016, ArXiv e-prints,
  arXiv:1603.08741

\bibitem[{{Haas} {et~al.}(2011){Haas}, {Chini}, {Ramolla}, {Pozo Nu{\~n}ez},
  {Westhues}, {Watermann}, {Hoffmeister}, \& {Murphy}}]{ha11}
{Haas}, M., {Chini}, R., {Ramolla}, M., {et~al.} 2011, \aap, 535, A73

\bibitem[{{H{\"o}nig} {et~al.}(2014){H{\"o}nig}, {Watson}, {Kishimoto}, \&
  {Hjorth}}]{ho14}
{H{\"o}nig}, S.~F., {Watson}, D., {Kishimoto}, M., \& {Hjorth}, J. 2014, \nat,
  515, 528

\bibitem[{{Jim{\'e}nez-Vicente} {et~al.}(2014){Jim{\'e}nez-Vicente},
  {Mediavilla}, {Kochanek}, {Mu{\~n}oz}, {Motta}, {Falco}, \&
  {Mosquera}}]{jjv14}
{Jim{\'e}nez-Vicente}, J., {Mediavilla}, E., {Kochanek}, C.~S., {et~al.} 2014,
  \apj, 783, 47

\bibitem[{{Kaastra} {et~al.}(2014){Kaastra}, {Kriss}, {Cappi}, {Mehdipour},
  {Petrucci}, {Steenbrugge}, {Arav}, {Behar}, {Bianchi}, {Boissay},
  {Branduardi-Raymont}, {Chamberlain}, {Costantini}, {Ely}, {Ebrero}, {Di
  Gesu}, {Harrison}, {Kaspi}, {Malzac}, {De Marco}, {Matt}, {Nandra},
  {Paltani}, {Person}, {Peterson}, {Pinto}, {Ponti}, {Nu{\~n}ez}, {De Rosa},
  {Seta}, {Ursini}, {de Vries}, {Walton}, \& {Whewell}}]{ka14}
{Kaastra}, J.~S., {Kriss}, G.~A., {Cappi}, M., {et~al.} 2014, Science, 345, 64

\bibitem[{{Kara} {et~al.}(2016){Kara}, {Alston}, {Fabian}, {Cackett}, {Uttley},
  {Reynolds}, \& {Zoghbi}}]{ka16}
{Kara}, E., {Alston}, W.~N., {Fabian}, A.~C., {et~al.} 2016, \mnras, 462, 511

\bibitem[{Komatsu {et~al.}(2011)Komatsu, Smith, Dunkley, Bennett, Gold,
  Hinshaw, Jarosik, Larson, Nolta, Page, Spergel, Halpern, Hill, Kogut, Limon,
  Meyer, Odegard, Tucker, Weiland, Wollack, \& Wright}]{ko11}
Komatsu, E., Smith, K.~M., Dunkley, J., {et~al.} 2011, The Astrophysical
  Journal Supplement Series, 192, 18

\bibitem[{{Korista} \& {Goad}(2001)}]{ko01}
{Korista}, K.~T., \& {Goad}, M.~R. 2001, \apj, 553, 695

\bibitem[{{Lira} {et~al.}(2015){Lira}, {Ar{\'e}valo}, {Uttley}, {McHardy}, \&
  {Videla}}]{li15}
{Lira}, P., {Ar{\'e}valo}, P., {Uttley}, P., {McHardy}, I.~M.~M., \& {Videla},
  L. 2015, \mnras, 454, 368

\bibitem[{{McHardy} {et~al.}(2016){McHardy}, {Connolly}, {Peterson}, {Bieryla},
  {Chand}, {Elvis}, {Emmanoulopoulos}, {Falco}, {Gandhi}, {Kaspi}, {Latham},
  {Lira}, {McCully}, {Netzer}, \& {Uemura}}]{mc16}
{McHardy}, I., {Connolly}, S., {Peterson}, B., {et~al.} 2016, ArXiv e-prints,
  arXiv:1601.00215

\bibitem[{{McHardy} {et~al.}(2014){McHardy}, {Cameron}, {Dwelly}, {Connolly},
  {Lira}, {Emmanoulopoulos}, {Gelbord}, {Breedt}, {Arevalo}, \&
  {Uttley}}]{mc14}
{McHardy}, I.~M., {Cameron}, D.~T., {Dwelly}, T., {et~al.} 2014, \mnras, 444,
  1469

\bibitem[{{Morgan} {et~al.}(2010){Morgan}, {Kochanek}, {Morgan}, \&
  {Falco}}]{mo10}
{Morgan}, C.~W., {Kochanek}, C.~S., {Morgan}, N.~D., \& {Falco}, E.~E. 2010,
  \apj, 712, 1129

\bibitem[{{Morgan} {et~al.}(2012){Morgan}, {Hainline}, {Chen}, {Tewes},
  {Kochanek}, {Dai}, {Kozlowski}, {Blackburne}, {Mosquera}, {Chartas},
  {Courbin}, \& {Meylan}}]{mo12}
{Morgan}, C.~W., {Hainline}, L.~J., {Chen}, B., {et~al.} 2012, apj, 756, 52

\bibitem[{{Mosquera} {et~al.}(2013){Mosquera}, {Kochanek}, {Chen}, {Dai},
  {Blackburne}, \& {Chartas}}]{mo13}
{Mosquera}, A.~M., {Kochanek}, C.~S., {Chen}, B., {et~al.} 2013, \apj, 769, 53

\bibitem[{{Narayan}(1996)}]{na96}
{Narayan}, R. 1996, \apj, 462, 136

\bibitem[{{Nealon} {et~al.}(2015){Nealon}, {Price}, \& {Nixon}}]{ne15}
{Nealon}, R., {Price}, D.~J., \& {Nixon}, C.~J. 2015, \mnras, 448, 1526

\bibitem[{{Pancoast} {et~al.}(2014){Pancoast}, {Brewer}, {Treu}, {Park},
  {Barth}, {Bentz}, \& {Woo}}]{pa14}
{Pancoast}, A., {Brewer}, B.~J., {Treu}, T., {et~al.} 2014, mnras, 445, 3073

\bibitem[{{Peterson}(1993)}]{pe93}
{Peterson}, B.~M. 1993, \pasp, 105, 247

\bibitem[{{Poindexter} \& {Kochanek}(2010)}]{Po10}
{Poindexter}, S., \& {Kochanek}, C.~S. 2010, \apj, 712, 668

\bibitem[{{Seaton}(1979)}]{se79}
{Seaton}, M.~J. 1979, \mnras, 187, 73P

\bibitem[{{Sergeev} {et~al.}(2007){Sergeev}, {Doroshenko}, {Dzyuba},
  {Peterson}, {Pogge}, \& {Pronik}}]{se07}
{Sergeev}, S.~G., {Doroshenko}, V.~T., {Dzyuba}, S.~A., {et~al.} 2007, \apj,
  668, 708

\bibitem[{{Sergeev} {et~al.}(2005){Sergeev}, {Doroshenko}, {Golubinskiy},
  {Merkulova}, \& {Sergeeva}}]{se05}
{Sergeev}, S.~G., {Doroshenko}, V.~T., {Golubinskiy}, Y.~V., {Merkulova},
  N.~I., \& {Sergeeva}, E.~A. 2005, \apj, 622, 129

\bibitem[{{Shakura} \& {Sunyaev}(1973)}]{ss73}
{Shakura}, N.~I., \& {Sunyaev}, R.~A. 1973, \aap, 24, 337

\bibitem[{{Shappee} {et~al.}(2014){Shappee}, {Prieto}, {Grupe}, {Kochanek},
  {Stanek}, {De Rosa}, {Mathur}, {Zu}, {Peterson}, {Pogge}, {Komossa}, {Im},
  {Jencson}, {Holoien}, {Basu}, {Beacom}, {Szczygie{\l}}, {Brimacombe},
  {Adams}, {Campillay}, {Choi}, {Contreras}, {Dietrich}, {Dubberley},
  {Elphick}, {Foale}, {Giustini}, {Gonzalez}, {Hawkins}, {Howell}, {Hsiao},
  {Koss}, {Leighly}, {Morrell}, {Mudd}, {Mullins}, {Nugent}, {Parrent},
  {Phillips}, {Pojmanski}, {Rosing}, {Ross}, {Sand}, {Terndrup}, {Valenti},
  {Walker}, \& {Yoon}}]{sh14}
{Shappee}, B.~J., {Prieto}, J.~L., {Grupe}, D., {et~al.} 2014, \apj, 788, 48

\bibitem[{{Starkey} {et~al.}(2016){Starkey}, {Horne}, \& {Villforth}}]{st15}
{Starkey}, D.~A., {Horne}, K., \& {Villforth}, C. 2016, \mnras, 456, 1960

\bibitem[{{Troyer} {et~al.}(2016){Troyer}, {Starkey}, {Cackett}, {Bentz},
  {Goad}, {Horne}, \& {Seals}}]{tr16}
{Troyer}, J., {Starkey}, D., {Cackett}, E.~M., {et~al.} 2016, \mnras, 456, 4040

\bibitem[{{Uttley} {et~al.}(2014){Uttley}, {Cackett}, {Fabian}, {Kara}, \&
  {Wilkins}}]{ut14}
{Uttley}, P., {Cackett}, E.~M., {Fabian}, A.~C., {Kara}, E., \& {Wilkins},
  D.~R. 2014, \aapr, 22, 72

\bibitem[{{Wambsganss}(2006)}]{wa06}
{Wambsganss}, J. 2006, ArXiv Astrophysics e-prints, astro-ph/0604278

\bibitem[{{Wanders} {et~al.}(1997){Wanders}, {Peterson}, {Alloin}, {Ayres},
  {Clavel}, {Crenshaw}, {Horne}, {Kriss}, {Krolik}, {Malkan}, {Netzer},
  {O'Brien}, {Reichert}, {Rodr{\'{\i}}guez-Pascual}, {Wamsteker}, {Alexander},
  {Anderson}, {Benitez}, {Bochkarev}, {Burenkov}, {Cheng}, {Collier},
  {Comastri}, {Dietrich}, {Dultzin-Hacyan}, {Espey}, {Filippenko}, {Gaskell},
  {George}, {Goad}, {Ho}, {Kaspi}, {Kollatschny}, {Korista}, {Laor},
  {MacAlpine}, {Mignoli}, {Morris}, {Nandra}, {Penton}, {Pogge}, {Ptak},
  {Rodr{\'{\i}}guez-Espinoza}, {Santos-Lle{\'o}}, {Shapovalova}, {Shull},
  {Snedden}, {Sparke}, {Stirpe}, {Sun}, {Turner}, {Ulrich}, {Wang}, {Wei},
  {Welsh}, {Xue}, \& {Zou}}]{wa97}
{Wanders}, I., {Peterson}, B.~M., {Alloin}, D., {et~al.} 1997, \apjs, 113, 69

\bibitem[{{White} \& {Peterson}(1994)}]{wh94}
{White}, R.~J., \& {Peterson}, B.~M. 1994, \pasp, 106, 879

\bibitem[{{Zu} {et~al.}(2011){Zu}, {Kochanek}, \& {Peterson}}]{zu11}
{Zu}, Y., {Kochanek}, C.~S., \& {Peterson}, B.~M. 2011, \apj, 735, 80

\end{thebibliography}
\bibliographystyle{apj}

\end{document}